%% file: main.tex
\documentclass[journal]{IEEEtran}
\usepackage{empheq,mathrsfs}

\usepackage{booktabs,mathtools,caption,subcaption}
\usepackage{amsmath,algorithmic,amsfonts}
\usepackage{xcolor,graphicx,tcolorbox,colortbl}
\usepackage{xspace,array}
\usepackage{url,textcomp}
\usepackage[hidelinks,bookmarks=false]{hyperref}
\usepackage{pifont}
\usepackage{multirow}
\usepackage{enumitem}
\usepackage{multicol}
\tcbuselibrary{theorems}
\newcommand*{\shadedbox}{%
 \tcboxmath[colframe=white, size=fbox, arc=0mm, boxrule=0.8pt]%
}
\newcommand*{\shadedboxb}{%
 \tcboxmath[colback = red!5!white, colframe=white, size=fbox, arc=0mm, boxrule=0.8pt]%
}
\newcommand*{\shadedboxc}{%
 \tcboxmath[colback = green!5!white, colframe=white, size=fbox, arc=0mm, boxrule=0.8pt]%
}
\newcommand*{\shadedboxd}{%
 \tcboxmath[colback = yellow!15!white, colframe=white, size=fbox, arc=0mm, boxrule=0.8pt]%
}
\newcommand*{\shadedboxe}{%
 \tcboxmath[colback = blue!10!white, colframe=white, size=fbox, arc=0mm, boxrule=0.8pt]%
}

\begin{document}\sloppy

\definecolor{darkbrown}{rgb}{.5,0,0}
\definecolor{dblue}{rgb}{0,0,.5}
\definecolor{lblue}{rgb}{.87,.87,.92}
\definecolor{vlblue}{rgb}{.98,.99,.99}

\newcommand*\widefbox[1]{\fbox{\hspace{2em}#1\hspace{2em}}}
\newcommand{\cmnt}[1]{\ignorespaces}
\newcommand{\T}{\mathcal{T}}
\newcommand{\E}{\mathbb{\bf E}}
\newcommand{\celsius}{$^\circ$C }
\newcommand{\spread}{f_{sp}}
\newcommand{\leakspread}{f_{leaksp}}
\newcommand{\utemp}{\mathcal{U}}
\newcommand{\la}{\mathcal{L}}
\newcommand{\pdyn}{\mathcal{P}}
\newcommand{\hankel}{\mathcal{H}}
\newcommand{\bessel}{\mathcal{J}}
\newcommand{\correl}{\mathcal{R}}
\newcommand{\fourier}{\mathcal{F}}
\newcommand{\fsilic}{f_{silic}}
\newcommand{\fdelta}{f_{\Delta}}
\newcommand{\ftrans}{f_{trans}}
\newcommand{\fetafunc}{ {f_\alpha} }
\newcommand{\finvfunc}{ {f_{inv}} }
\newcommand{\fname}{\textit{VarSim}\xspace}
\newcommand{\kignore}{1.5\%}
\newcommand{\speedup}{200X\xspace}
\newcommand{\error}{4\%}
\newcommand{\Pleak}{P_{leak_0} }
\newcommand{\red}[1]{\textcolor{red}{#1}}
\newcommand{\gray}[1]{\textcolor{darkgray}{#1}}
\newcommand{\dblue}[1]{\textcolor{dblue}{#1}}
\newcommand{\br}[1]{\textcolor{darkbrown}{#1}}
\newcommand{\one}{\ding{182}\xspace}%
\newcommand{\two}{\ding{183}\xspace}%
\newcommand{\three}{\ding{184}\xspace}%
\newcommand{\four}{\ding{185}\xspace}%
\newcommand{\five}{\ding{186}\xspace}%
\newcommand{\six}{\ding{187}\xspace}%

\title{VarSim: A Fast Process Variation-aware Thermal Modeling Methodology Using Green's Functions
}

\author{Hameedah Sultan
        and~Smruti~R~Sarangi
\IEEEcompsocitemizethanks{\IEEEcompsocthanksitem Hameedah Sultan is with the School of Information Technology, Indian Institute of Technology, Delhi, India, 110016.\protect\\
E-mail: hameedah.sultan@gmail.com
\IEEEcompsocthanksitem Smruti R Sarangi is with the Department of Computer Science and Engineering, joint faculty in the Electrical Engineering department, Indian Institute of Technology, Delhi, 110016.\protect\\
E-mail: srsarangi@cse.iitd.ac.in}
}

\IEEEtitleabstractindextext{%
\input{abstract}

\begin{IEEEkeywords}
thermal simulation, Hankel transform, process variation, temperature dependent conductivity, leakage  power,  Green's functions
\end{IEEEkeywords}}

\maketitle

\IEEEdisplaynontitleabstractindextext

\input{introduction}

\input{background}

\input{related}
\input{methodology}

\input{evaluation}

\input{conc}

\input{main.bbl}

\input{bios}

\appendix
\input{appendix}
\end{document}

%% file: abstract.tex
\begin{abstract}
Despite temperature rise being a first-order design constraint, traditional thermal estimation techniques have severe limitations in modeling critical aspects affecting the temperature in modern-day chips. 
Existing thermal modeling techniques often ignore the effects of parameter variation, which can lead to significant errors. Such methods also ignore the dependence of conductivity on temperature and its variation. Leakage power is also incorporated inadequately by state-of-the-art techniques. Thermal modeling is a process that has to be repeated at least thousands of times in the design cycle, and hence speed is of utmost importance.

To overcome these limitations, we propose \fname, an ultrafast thermal simulator based on Green's functions. 
Green's functions have been shown to be faster than the traditional finite difference and finite element-based approaches but have rarely been employed in thermal modeling. 
Hence we propose a new Green's function-based method to capture the effects of leakage power as well as process variation analytically. We
provide a closed-form solution for the Green's function considering the effects of variation on the process, temperature, and thermal conductivity. In addition, we propose a novel way of dealing with the anisotropicity introduced by process variation by splitting the Green's functions into shift-variant and shift-invariant components. Since our solutions are analytical expressions, we were able to obtain speedups that were several orders of magnitude over and above state-of-the-art proposals with a mean absolute error
limited to 4\% for a wide range of test cases. Furthermore, our method accurately captures the steady-state as well as  the transient variation in temperature.
\end{abstract}

%% file: introduction.tex
\section{Introduction}
\label{sec:intro}
The demand for high-performance computing as well as applications in fields such as machine learning, big data analytics, IoT, and edge computing has led to increased power densities in modern-day chips.
The resultant temperature rise has several harmful effects that include an increase in leakage power and a disproportionate 
decrease in reliability. 
Hence, thermal simulation is now one of the most critical
steps in the overall semiconductor design flow. It is typically a long and time-consuming process, that has to be repeated several times for a multitude of use cases in the design cycle.

To make matters worse, process variation has increasingly been leading to large deviations in 
electrical and thermal parameters of transistors, thereby leading to a high degree of unpredictability in key circuit parameters such as the timing delay and leakage power consumption. 
With ongoing device scaling, handling and mitigating process variation continues to become increasingly critical. Process variation affects all major architectural design decisions. 

Sadly, thermal modeling in chips with process variation is extremely complex and slow; till date no fast and efficient solutions have been 
proposed; researchers still rely on traditional Finite Element Method (FEM) and Finite Difference Method (FDM) analysis.

There is however a {\bf strong need} for fast thermal simulation methods in this space.
Many architectural techniques have been proposed with the aim of mitigating the adverse
effects of process variation such as functional unit level body biasing and retiming. However, to effectively incorporate such schemes, an accurate estimate of the impact of variation is needed. This requires extensive thermal simulations for a wide range of power plans.
Similarly, while floorplanning or cell placement, a large number of optimization strategies need to be quickly evaluated for a range of process variation scenarios~\cite{elseviermcm}. Leakage power has a strong temperature dependence and is heavily influenced by process variation as well. The chip temperature itself is dependent on leakage power, resulting in a cyclic effect. Thus leakage has a significant impact on the temperature of modern-day chips, often contributing to half of the total temperature rise. 
\textbf{Thus, a proper design optimization requires performing a thorough thermal evaluation through
simulations on chips having a significant range of process, conductivity and temperature variation, while correctly incorporating leakage power.} Hence, there is a need for a \textit{fast} thermal simulator in this space. A disclaimer is due -- at different stages of the
design process the designer has different degrees of information. Nevertheless, there is still a need for ultra-fast thermal simulation
because designers always seek the most productive design choices with the information that they have at that stage.
For example, at the architectural level, in the product planning stage, just a broad idea of the power consumption is available. Hence, the designer uses high level core power and variation models here, which still need to factor in the effects of temperature. After RTL signoff, placement and routing, synthesis and layout, progressively more accurate power numbers are available at each stage, and any change introduced at any level requires further modeling updates to guide the optimal design choices. This exercise needs to be done for every DCVS corner. After tapeout in the post silicon stage, the exact dynamic power numbers and a concrete idea of the process variation is available. 
Thus at every stage of the design process, each simulation produces crisp, exact data, and thousands of such simulations are run with different power dissipation (variation) values for a range of use cases. 
This Monte Carlo simulation yields a distribution that helps determine the final chip design. 
This is the standard practice widely used across the semiconductor industry.
 
Unfortunately, existing {\bf architectural thermal simulators do not consider the effects of process variation}. Ignoring process variation could lead to failure of the device after fabrication~\cite{sapatnekarvar}. Additionally, most simulators fail to factor in the temperature dependence of conductivity, leading to significant errors in thermal estimation. Prior work has shown that ignoring the temperature dependence of conductivity can result in an error of up to 5\celsius~\cite{isac}. Furthermore, traditional thermal simulators are based on the costly finite element and finite difference methods, making them slow and limiting the scope of design space exploration. 
On top of that, most existing thermal simulators consider the effects of leakage power by iterating through the leakage-temperature feedback loop, increasing their runtime several times. Since leakage power is indispensable in modern-day chips, it is essential for today's thermal simulators to naturally consider the effects of leakage power as part of the core modeling methodology and avoid iterative computations. 

Consequently, fast thermal estimation that takes variability into account has hitherto remained an open problem.
To solve this problem, we propose a thermal simulation methodology, \fname, in this paper. \textbf{\fname is a novel Green's function-based analytical thermal simulator, that inherently captures the effects of both process variation as well as leakage power, without running costly iterations.} Our main contributions can be summarized as follows:

\begin{enumerate}[wide, labelwidth=!, labelindent=0pt]
\item \fname considers the impact of process variation as well as the temperature dependence of conductivity. To the extent of our knowledge, no existing technique has done this. 

\item Our method is based on Green's functions (impulse response of a power source), which are known to be very fast methods~\cite{3dsim,sapatnekar}. However, Green's functions in thermal estimation rely on the shift-invariance of the impulse response, which ceases to be true when process variation is considered. We propose a novel way of overcoming this limitation by splitting the Green's function into a shift-invariant component and a shift-variant component.

\item We mathematically derive a novel, modified leakage-aware Green's function that incorporates in itself the impact of temperature-dependent conductivity and leakage power. This modified Green's function is directly used at runtime with the dynamic power profile to obtain an accurate estimate of the full-chip thermal profile considering all the desired effects. 
Since our method is analytical, it is also extremely fast. 
\item Green’s function-based methods have great potential, since their
accuracy is not dependent on the grid size, making them faster. However,
their applicability is limited by a lack of solutions for modern
EDA problems. Thus our work contributes to an ecosystem, where
more researchers can propose faster Green’s function-based solutions
for newer problems.
\end{enumerate}

Using an analytical Green's function-based approach, we obtain 
a several orders of 
magnitude speedup over state-of-the-art approaches, while keeping the maximum error within \error.
\textbf{Thus by leveraging the underlying physics of heat transfer and process variation, we are able to increase the accuracy of thermal simulations by considering all the critical temperature-affecting phenomena, while simultaneously achieving a very high simulation speed.} The speed advantage becomes even more pronounced when thermal simulations need to be repeated thousands of times, as is done in a typical design cycle. 

The rest of the paper is organized as follows.
We describe some background information for our work in Section~\ref{sec:backgnd}. The current state-of-the-art literature
is described in Section~\ref{sec:related}. We describe our modeling methodology in Section~\ref{sec:method}. Section~\ref{sec:eval} describes the evaluation of our proposed method and the corresponding results. We finally conclude in Section~\ref{sec:conc}.

%% file: background.tex
\section{Theoretical Preliminaries}
\label{sec:backgnd}

\subsection{Fourier Equation}
The temperature distribution in a body is governed by the Fourier heat equation~\cite{fourier}, described mathematically as:
\begin{equation}
\label{eqn:fourier}
 \nabla.\left(\kappa\nabla T\right) + \dot{q} = \rho C_v \frac{\partial T}{\partial t},
\end{equation}
where $\rho$ is the material density, $T$ is the temperature, $C_v$ is the volumetric specific heat, $\kappa$ is the thermal 
conductivity, and $\dot{q}$
is the rate of heat energy generation inside the volume.
 However, this equation is too complex to be solved analytically in the general case. Hence most chip-level thermal simulators use numerical methods such as the finite difference and the finite element methods to arrive at an approximate solution quickly.

\subsection{Green's Functions}
An alternative approach to obtain the temperature profile is based on the impulse response of the chip (or the 
Green's function~\cite{greenintro}). This impulse response is obtained by applying a unit power source to the center of the chip and getting the  corresponding temperature rise. {The Green's function is essentially the heat spread function, $f_{sp}$, of a point power source.}
 This Green's function is then convolved with the power profile to obtain the full-chip temperature profile. This approach is analytical, and 
much faster than finite difference or finite element-based approaches since the entire heat transfer path is not modeled, rather 
only the power dissipating layers and the boundary conditions are considered~\cite{lightsim,3dsim, sapatnekar}.
This is because the accuracy of the finite element method is dependent on the grid size, and reducing the grid size comes at the expense of lost accuracy. However, because of the analytical nature of the Green's function, the grid size here merely determines the output resolution and the accuracy of the method is independent of the grid size.

Using the impulse response (Green's function), the complete full-chip temperature profile can be calculated as~\cite{powerblur2014}:
\begin{equation}
\label{eqn:Green}
T = f_{sp} \star P
\end{equation}
where $P$ is the power dissipation profile, $f_{sp}$ is the Green's function, and $\star$ is the convolution operator.

{The Green's function is radially symmetric and can be further decomposed into a rapidly decaying part $f_{silic}$, and a constant representing heat redistributed through the heat spreader~\cite{lightsim}.}

Interested readers may read more about the Green's function in~\cite{lightsim, 3dsim, sapatnekar}.

\subsection{Process and Temperature Variation}
The manufacturing of an IC involves a large number of steps or processes that are imperfect in nature. As a result, the properties of a manufactured chip often differ from its nominal values. The device dimensions have reached the scale of tens of atoms in modern-day chips. As a result, the impact of variation has become much more prominent. 

The parameters affected by variation include the
oxide thickness, threshold voltage, gate width, and channel length.
The variation in these parameters is classified as: \textit{wafer-to-wafer}, \textit{die-to-die} and \textit{within-die} variations. 
The first two effects (collectively known as \textit{inter-die} variation) uniformly affect all regions of a given die. They used to have a larger significance in older 
technology generations; they can be mitigated easily by relatively simple methods such as \textit{frequency binning}. These typically cause a constant shift in the mean value of a parameter across all the devices on a die. 

For newer 
technology generations, within-die variation dominates and requires more complex management 
strategies~\cite{mittal}. This type of variation leads to deviations in the electrical and thermal properties of the chip on the same die. Within-die variation is further classified as:

\begin{enumerate}[wide, labelwidth=!, labelindent=0pt]
\item \textbf{Systematic variations:} These are introduced because of lithographic aberrations and diffraction or chemical-mechanical polishing/planarization (CMP) effects. Systematic variation results in proximate regions on the die having similar values of parameters. It is modeled by a multivariate Gaussian 
distribution~\cite{varius} having a spherical correlation~\ref{fig:pleakcorr}.   

\begin{figure}
\centering
\includegraphics[trim={0 0 0 0},clip,width= 0.9 \columnwidth]{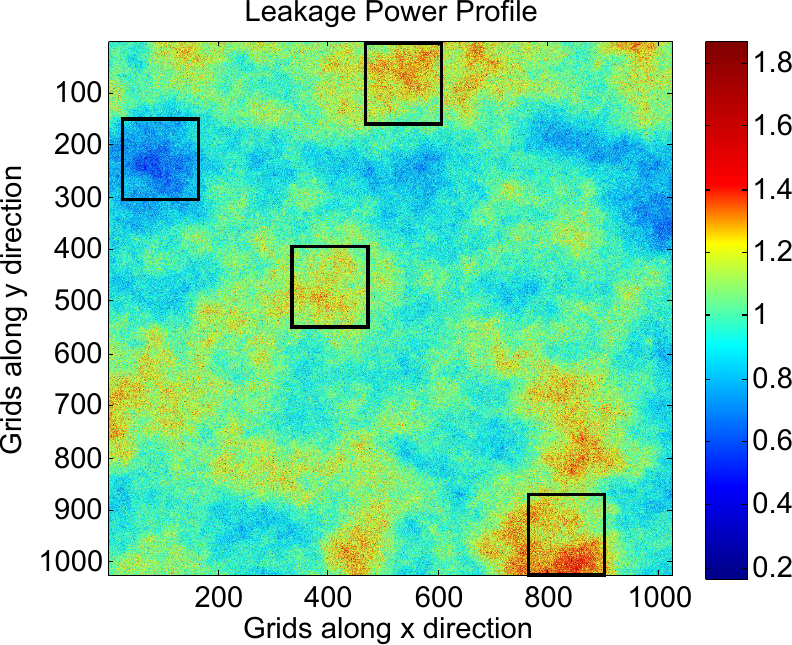}
\caption{Leakage power in the presence of variability. Note the spatial correlation 
in the leakage power values {(high concentration of similar values in the bounding boxes)}. \label{fig:pleakcorr}}
\end{figure}

\item \textbf{Random variations:} These are caused by random dopant fluctuations 
(RDF) and line edge roughness; they are together modeled as a 
zero-mean Gaussian random variable. These variations do not exhibit any spatial correlation.
\end{enumerate}

There are two variables in the heat equation that are strongly affected by parameter 
variation: leakage power and thermal
conductivity. We discuss these next.

\subsection{Leakage power}
The variability in leakage power arises because of both systematic and random variations. However, it is well-known~\cite{skadronvar} 
that the effects of random variation tend to get averaged out at the architectural level when considering temperature. 
We have also observed the same in our experiments.

\label{sec:leakagedef}
The subthreshold leakage current, $I_{leak}$ is given by Equation~\ref{eqn:bsim}.
\begin{equation}
\label{eqn:bsim}
 I_{leak} \propto v_{T}^{2} * e^{\frac{V_{GS}-V_{th}-V_{off}}
{\eta * v_{T}}} (1-e^{\frac{-V_{DS}}{v_{T}}})
\end{equation}

where, $v_T$ is the thermal voltage ($kT/q$), $V_{th}$ is the threshold voltage, $V_{off}$ is the offset voltage in the
sub-threshold region and $\eta$ is a constant. Because of variability, the oxide thickness and gate length 
change, which result in a change in the threshold voltage. 
The temperature 
dependence of $I_{leak}$ can be 
modeled with a reasonable accuracy using a linear equation~\cite{lightsim,3dsim, liu}. Equation~\ref{eqn:bsim} then becomes:

\begin{equation}
\label{eqn:Ileakmod}
 I_{leak} \propto (1+\beta \Delta T)e^{\beta_{ L}\Delta L + \beta_{ t_{ox}}\Delta t_{ox}}
\end{equation}

where $\beta$ represents the change in leakage power with temperature, $\beta_{ t_{ox}}$ is a constant 
representing the variability in the oxide thickness $t_{ox}$ and $\beta_{L}$ 
represents the variability in the gate 
length, $L$.
The corresponding leakage power is given by:
\begin{equation}
\label{eqn:Pleak}
 P_{leak} = (1+\beta \Delta T) P_{leak_0},
\end{equation}
where $\Pleak$ is the leakage power at ambient temperature after considering the impact of variability.
For improved accuracy, we can use a piece-wise linear leakage model, which provides an accuracy of over 
99\%~\cite{tempsurvey}.

\subsection{Conductivity of Silicon}
In addition to considering the impact of variation on leakage power, we also consider the impact of variability on the conductivity of silicon. This is because in accordance with the Fourier's law, the temperature profile is impacted by both the power consumed by the chip as well as the conductivity of the chip. 
Random dopant fluctuations (RDF) cause a variation in the doping profile, resulting in variations in the conductivity of the material as well. 
To model this variation in conductivity, we consider a Gaussian random variable, $K$. The range of variation in the doping profiles because of RDF is obtained from the literature~\cite{leungvar}. The range of conductivity values of silicon for these dopant densities is obtained from the literature~\cite{burzo}. Using these, the variance in the conductivity of silicon is then obtained.

In addition to the random variation in conductivity, the conductivity of silicon also depends on temperature. The conductivity of silicon varies with temperature according to the following equation:
\begin{equation}
\label{eqn:condsi}
\kappa = k_{0} \left({T \over 300}\right)^{-\eta}, 
\end{equation}
where $k_{0}$ is the conductivity of silicon at 300K, $T$ is the temperature in Kelvin, and $\eta$ is a material-dependent constant. 
As the chip gets heated, the conductivity of silicon decreases which further affects the temperature profile.

\subsection{Transforms used in this Paper}
Thermal problems are often easier to solve in the transform domain. We use two types of transforms in this work -- the \textit{Fourier transform}  and the \textit{Hankel transform}.

\subsubsection{Fourier Transform} 
The Fourier transform decomposes a signal from the spatial domain and brings it into the frequency domain. 
The result is a complex function, the magnitude of which represents the amount of each frequency present in the signal.
In the present work, we make use of the 2-dimensional Fourier transform, which is given by:
\begin{equation}
\begin{split}
\label{eqn:fourierdef}
F(u,v) = \fourier(f(x,y)) = \int^{\infty}_{-\infty} \int^{\infty}_{-\infty} f(x,y) e^{-j 2 \pi(ux+vy)}dx dy\\
\end{split}
\end{equation}
where $u,v$ are the Fourier frequency domain variables, $x,y$ are the spatial domain variables, and $f(x,y)$ is the spatial domain signal being transformed into the frequency domain.

\subsubsection{Hankel Transform}
The Hankel transform is equivalent to the 
2-D Fourier transform of a radially symmetric function. It uses the Bessel function as its basis. The Hankel transform is
defined as: 
\begin{equation}
\label{eqn:hankeldef}
\hankel(f(r)) = H(s) = \int_0^\infty f(r) \bessel_0(sr)r dr,
\end{equation}
where $\bessel_0$ is the Bessel function of the first kind of order 0, and $\hankel$ denotes the Hankel transform operator.

%% file: related.tex
\section{Related Work}
\label{sec:related}
{Thermal modeling has been a focus area of the EDA industry in the last two decades, and hence researchers have extensively worked on various aspects of this problem, such as 2D and 3D ICs~\cite{hotspot, pod, systemc,deepoheat,pathania3dttp}, smartphones and other mobile devices, thermal-aware DNN accelerators~\cite{coskun2023} .
 
Process variation per se has been widely studied along with techniques for mitigating its pernicious effects. 
However, very rarely has the effect of process variation on temperature been looked at.
The works that do consider the effects of process variation on temperature, often do so by neglecting the temperature dependence of leakage power~\cite{jaffari}. We demonstrate that considering the effects of process variation on leakage power, but neglecting its temperature dependence may result in a 4 to 6\celsius error.

Prior works have also established the importance of modeling the temperature-dependence of the conductivity of silicon as well, and proposed methods to tackle the problem. However, such approaches do not consider process variation. Since both of these effects have never been considered simultaneously before, we look at each of these effects in related work separately.

\subsection{Effects of Process Variation on Temperature}
Varipower~\cite{varipower} models power variability at the architectural functional unit level by performing circuit-level Monte Carlo simulations incorporating parameter variation. However, it does not model the effects of variability on temperature. 

Humenay et al.~\cite{skadronvar} recognized the challenges imposed by systematic variation in ensuring homogeneous performance across cores. They demonstrated a large variation in power, temperature and performance across cores because of core-to-core systematic variation.

Jaffari and Anis~\cite{jaffari} statistically calculated the expected value of temperature considering the impact of variability. They first obtain the leakage-converged temperature iteratively without considering variation and then statistically compute the effect of parameter variation. They use their technique to iteratively update the computed power and temperature to estimate the full-chip power   and the probability density function of the temperature distribution. However, a significant limitation of their technique is that it is iterative, 
making it extremely slow ($\approx 158s$ for a $50\times50$ grid), 121X slower than 
\fname. 
Juan et al. \cite{juan} use a linear regression-based model to train and predict the maximum temperature in a 
3D IC in the presence of variability. They use measured values of leakage power for training. They demonstrate that 3D ICs are much more susceptible to variation, as compared to 2D ICs. However, 
learning-based methods are very sensitive to input data and do not generalize well when test conditions change. Additionally, their method captures the maximum temperature only, and not the complete thermal profile.

Shafique et al.~\cite{henkelvar} propose a variability-aware dark silicon management technique in which the cores to be throttled are determined on their workload patterns while accounting for the temperature map and variability. They propose a complex heuristic for predicting the temperature profile considering process variation. They superpose the impact of variability-affected leakage power on the estimated temperature map for a given thread-to-core mapping.
This heuristic attempts to manually approximate the underlying logic that is captured well   in our modified Green's function based approach. Our proposed method achieves this in a precise and efficient mathematical manner using the convolution operations. In comparison to our method, their technique is inexact, unnecessarily complex and slow. 

{Srinivasa et al.~\cite{pvsmartphone} demonstrate using measurements that because of process variation, smartphones of the same model may show a variation of upto 10-12\% in energy and performance.}

\subsection{Modeling the Temperature Dependence of Conductivity}
Yang et al.~\cite{isac} propose a temporally and spatially adaptive thermal analysis technique that accounts for the temperature dependence of conductivity. However, they do not consider leakage power.

Li et al.~\cite{isacvar} calculate the leakage power in the presence of variation, and use this as an input to the ISAC thermal modeling tool~\cite{isac}. They update leakage power in each time step and   iteratively proceed towards convergence. Ultimately, they use this augmented tool to study process variation in network-on-chips. This approach is iterative, and would require a large number of iterations to get to accurate variation-aware leakage-converged temperature values. For the sake of
comparison, we implement a similar approach using the HotSpot thermal modeling tool~\cite{hotspot5}, and demonstrate several orders of magnitude speedup using our method over such approaches.

Ziabari et al.~\cite{ziabari} consider the temperature-dependence of conductivity by using a lookup table to store Green's functions with different conductivities. At runtime, they iteratively update the Green's function until the temperature profile converges. They, however, do not model leakage or process variation.
In comparison, our approach encompasses the effects of leakage power, variability in leakage, and temperature-dependent conductivity analytically without requiring costly iterations.

Köroğlu and Pop~\cite{insulators} propose high thermal conductivity insulators for effective thermal management in 3D ICs.

There is an extensive body of work in fast transient thermal modeling as well (~\cite{MLtran,comet}). These works focus on fast runtime thermal estimation, 2.5 and 3D stacked packages, and thermal models for TSVs but do not take process variation into account.

{He et al.~\cite{he} propose a novel polynomial chaos for modeling uncertainty at the architectural level using mixed integer programming.}

{Chittamuru et al.~\cite{pvnoc} demonstrate the sensitivity of photonic networks-on-Chip (NoCs) to thermal and process variation and propose a robust framework to overcome its impact on reliability.
}

\textbf{However, process variation, along with the variation of conductivity with temperature, has 
never been considered before in a leakage-aware thermal simulation tool. }

%% file: methodology.tex
\section{Thermal Estimation Considering Variability}
\label{sec:method}

\subsection{Overview}

\begin{figure*}[!htbp]
\centering        
\makebox[\columnwidth]{%
\begin{minipage}{\columnwidth}
		\includegraphics[width=\columnwidth]{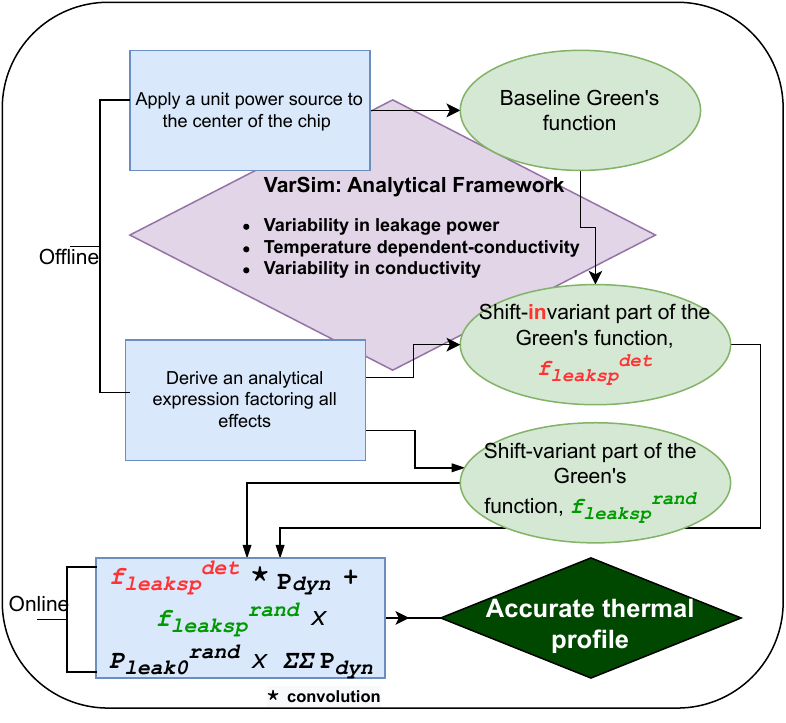}
		\subcaption{Overview of our approach\label{fig:flow}} 
	\end{minipage}%
		\begin{minipage}{\columnwidth}
		\includegraphics[width=1.2\columnwidth]{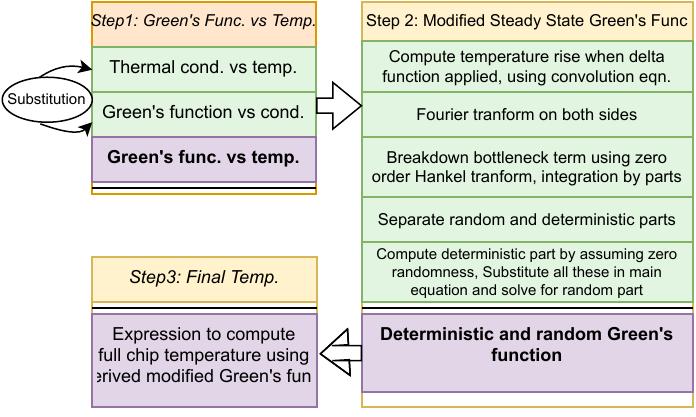}
		\subcaption{Overview of the mathematical derivation \label{fig:mathflow}}
	\end{minipage}
	}
        \caption{A high-level overview}\label{fig:fulloverview}
        \end{figure*}

Our approach fundamentally uses the Green's functions to compute the full-chip thermal profile. These can either be theoretically calculated, or empirically obtained by applying a unit impulse power source to the center of the chip, and measuring the corresponding temperature rise.
To consider the feedback effect of leakage power on temperature, Sarangi et al.~\cite{lightsim} derived a leakage-aware Green's function that captures this temperature-dependence of leakage power.
However, there are two effects that have not been considered in the use of these functions for full-chip temperature estimation: 
\begin{enumerate}[wide, labelindent=0pt]
\item The first effect is that the conductivity of silicon is also temperature-dependent, and has a non-negligible effect on temperature.
To capture this effect, we derive a novel modified Green's function, that not only captures the temperature-dependence of leakage power, but also captures the temperature-dependence of conductivity.

\item  The second unmodeled effect is process variation. Process variation poses multiple challenges in the use of Green's function-based methods.
The first challenge is in deriving the Green's function itself considering the effects of process variation as well.
{In a previous version of this work~\cite{varsim}, we had used the expected value of the leakage power map in place of the baseline leakage power map, to simplify the derivation of the modified temperature-dependent Green's function (note that the Green's function itself had captured the baseline variability, and this was not approximated; the approximation was in computing the temperature-dependent part of the leakage power). 
In the current work, we get rid of this approximation, and work out the complete expression.}

The second challenge is in computing the full-chip thermal profile using the derived Green's function. Temperature estimation using the Green's function relies on its \textit{shift-invariance}, which means that a power applied to any location of the chip will cause the same temperature rise, irrespective of the location of the power source on the chip (aside from the boundary effects, which need to be handled separately). This stops being true when process variation is considered. Hence, we split the Green's function into random and deterministic components. Then, we propose a novel way of combining the two components to generate the full-chip temperature profile. The deterministic component of the full-chip temperature profile is computed the regular way, using a convolution operation. The random component is location-dependent and hence is computed using the Hadamard product.

\end{enumerate}

We provide an overview of our proposed approach in Figure~\ref{fig:flow}. 
Table~\ref{tab:glossary} lists the terms used in our derivations.

Thus, our contributions can be summarized as follows:
\begin{enumerate}[wide, labelwidth=!, labelindent=0pt]

\item [\one] We derive an analytical expression for a novel leakage aware Green's function accounting for 
variability and 
the dependence 
of leakage power and conductivity on temperature. This Green's function has two components -- a shift-invariant component that accounts for the temperature dependence of conductivity and leakage power, and a shift-variant component that accounts for process variation.
We derive separate modified Green's functions for the steady-state (Section~\ref{sec:steady}) and the transient case (Section~\ref{sec:tran}). Furthermore, we describe the use of the transient Green's function to compute the temperature profile for a time-varying source in Section~\ref{sec:truetran}.

\item [\two] We calculate the full-chip variability-aware thermal profile in the presence of leakage as 
well as dynamic power by convolving the shift-invariant modified Green's functions with the dynamic power map and multiplying the shift-varying Green's functions with the baseline leakage power map considering process variation.

\end{enumerate}

\begin{table}[!hbtp]
\centering
\caption{Glossary \label{tab:glossary}}
	\begin{tabular}{c p{7cm}}
	\toprule
	\bf{Symbol} & \bf{Meaning} \\
	\midrule
	$\Pleak$ 	& Leakage power at ambient temperature considering variability\\
	$\beta$	 	& Temperature dependence of leakage power\\
	$\alpha$ 	& Temperature dependence of conductivity\\
	$\kappa$ 	& Conductivity of silicon\\
	$T$		 	& Temperature\\
	$\T$		& Temperature rise above ambient temperature\\
	$f_{sp_0}$	& Green's function without considering leakage and temperature-dependent conductivity $= f_{silic_0} + \phi $\\
	$\E(X)$		& Expected value of $X$\\
	$\fourier$	& Fourier transform operator \\
	$\hankel$		& Hankel transform operator\\
	$x,y$			& Spatial coordinates\\
	$u,v$			& Fourier frequency domain variables\\	
	$h$			& Hankel variable\\	
	$t$			& Time\\
	$C$			& Thermal capacitance\\
	$f_{leaksp}^k$ & Leakage aware Green's functions considering temperature dependence of conductivity\\
	$k_0'$ 		& Nominal conductivity of silicon at ambient temperature\\
	\bottomrule
	\end{tabular}
\end{table}

\subsection{Variation of the Thermal Conductivity with Temperature}
The conductivity of silicon varies with temperature according to the following relation~\cite{yuK}:
\begin{equation}
\kappa = k_{0} \left({T \over 300}\right)^{-\eta}, 
\label{eqn:kvsT}
\end{equation}
where $k_{0}$ is the conductivity of silicon at 300K, $T$ is the temperature in Kelvin, and $\eta$ is a material-dependent constant. 
In the operating range of ICs ($40 - 100$\celsius), we can linearize Equation~\ref{eqn:kvsT}:
\begin{equation}
\label{eqn:linkvsT}
\kappa(T) = {k_{0}}'(1-c\Delta T), 
\end{equation}
where $k_0'$ is the nominal conductivity of silicon at the ambient temperature, and $c$ is a constant. 

Next, we vary the conductivity of silicon and observe the change in the Green's function $f_{sp}$. We then obtain an empirical relation between the Green's function and conductivity (using HotSpot~\cite{hotspot6}):

\begin{equation}
\label{eqn:fspvsk}
f_{sp}(\kappa) = f_{sp_0}(1 - c'(\kappa(T) - k_0')), 
\end{equation}

where, $f_{sp_0}$ is the baseline Green's function when the temperature dependence of conductivity is not considered (the variation in conductivity because of random dopant fluctuations is captured in $f_{sp_0}$), and $c'$ is a constant.
Combining Equations~\ref{eqn:linkvsT} and \ref{eqn:fspvsk}, we obtain a relation for the Green's function that captures the dependence of conductivity on temperature.
\begin{equation}
\label{eqn:fspvsT}
f_{sp}(T) = f_{sp_0}(1 + \alpha\Delta T),
\end{equation}
where $\alpha$ is a constant that captures how much the Green's function varies because of a change in conductivity with temperature. 

\subsection{Modified Green's Functions for the Steady-State}
\label{sec:steady}
\subsubsection{Formulating the equation for the modified Green's function considering all effects}
Next, we derive the Green's function considering temperature-dependent conductivity, as well as temperature-dependent leakage power. We start with
an approach that is similar to that adopted by Sarangi et al.~\cite{lightsim} while also incorporating the effects of process variation and
the temperature dependence of conductivity.

The total power consumption ($P$) is the sum of the dynamic power ($P_{dyn}$) and the leakage power ($P_{leak}$). Using 
Equation~\ref{eqn:Pleak}, we get:
\begin{equation}
\label{eqn:Pt}
P = P_{dyn} + \Pleak(1 + \beta \Delta T).
\end{equation}

We assume that the dynamic power dissipation in the chip is initially zero. This implies that any temperature rise above the ambient temperature is because of leakage. From Equation~\ref{eqn:Green}, we have (subscript $_0$ refers to the initial state):
\begin{equation}
\label{eqn:T0}
T_0 = f_{sp_0} \star \Pleak
\end{equation}

Now, let us apply a unit impulse (Dirac delta function) dynamic power source to the center of the chip. Using the results in Equations~\ref{eqn:Green},
 \ref{eqn:fspvsT}, and \ref{eqn:Pt}  we get the updated temperature, $T_f$:
\begin{equation}
\label{eqn:Tf}
T_f = f_{sp_0}(1 + \alpha \Delta T) \star \left(\delta(x,y) +  \Pleak(1 + \beta \Delta T)\right)
\end{equation}
where $x,y$ are the spatial coordinates.

Next, we find out the increase in temperature because of the unit power source considering process variation and the temperature-dependent effects. We also use the property that the convolution of a function and a delta function is the function itself. We then arrive at:
\begin{equation}
\begin{split}
\label{eqn:Tsub}
\T &= T_f - T_0 \\
 &= f_{sp_0}(1 + \alpha\T)  + f_{sp_0}(1 + \alpha\T) \star \Pleak(1 + \beta \T) \\& 
 - f_{sp_0} \star \Pleak
\end{split}
\end{equation}

We need to solve for the temperature rise, $\T$, here. To convert the convolution operation into multiplication, {we compute the Fourier transform on both sides and} apply the property that the Fourier transform of the convolution of two functions is equal to the product of their individual Fourier transforms.
We compute the Fourier transform of both sides of Equation~\ref{eqn:Tsub} to 
arrive at 
Equation~\ref{eqn:FTf}. 
\begin{equation}
\begin{split}
\label{eqn:FTf}
\fourier(\T) &= \left( \fourier(f_{sp_0}) + \alpha \fourier(f_{sp_0}\T)\right) + \left(\fourier(f_{sp_0}) + \alpha \fourier(f_{sp_0}\T)\right) \times\\
			  &\left(\fourier(\Pleak) + \beta\fourier(\Pleak \T)\right) - \fourier(f_{sp_0}) \fourier(\Pleak)\\		  
			 &=  \underbrace{\fourier(f_{sp_0})}_{I} + \shadedboxb{\underbrace{\alpha \fourier(f_{sp_0}\T)}_{II}} +  \shadedboxc{\underbrace{\beta \fourier(f_{sp_0})\fourier({\Pleak \T})}_{III}}\\ & {+} 
\shadedboxd{\underbrace{\alpha \fourier(f_{sp_0}\T)\fourier(\Pleak)}_{IV}}
 + \shadedboxe{\underbrace{\alpha \beta\fourier(\Pleak \T)\fourier(f_{sp_0}\T)}_{V}}
\end{split}
\end{equation}

In Equation~\ref{eqn:FTf}, {$\fourier$ is the Fourier transform operator,} term $I$ is the baseline Green's function, term $II$ corresponds to the increase in 
temperature because of the temperature-dependence of conductivity, term $III$ is the increase in temperature because 
of the temperature-dependence of leakage power, term $IV$ corresponds to the compounded 
effect of the baseline leakage 
power and the temperature-dependence of the conductivity, and term $V$ is the 
increase in temperature because of the compounded effects of  temperature-dependent conductivity and leakage. The last term here is small because each of the temperature-dependent variables (conductivity/leakage power) by itself do  not cause a large enough change in the other variable to result in a large temperature change. Hence, we 
neglect this term.

\subsubsection{Reducing the bottleneck term (Term II)}

The most difficult term to compute in the above equation is $\fourier(f_{sp_0}\T)$. Let $G(u,v) =  \fourier(f_{sp_0}\T)$. 
For this we make use of Lemma I described below:
\subsubsection*{Lemma I}: $G(u,v) =  \fourier(f_{sp_0}\T) = \fourier (\T) g_{sp_0}$, where $g_{sp_0} = \left( f_{sp_0} - \kappa + f_{sp_0}(0,0)\right)$

Interested readers may refer to the proof of Lemma 1 in the appendix. 	   

\noindent Using Lemma I in Equation~\ref{eqn:FTf}, we have:

\begin{equation}
\begin{split}
\label{eqn:expanded1}
 \fourier(\T) &= \fourier(f_{sp_0}) + \alpha g_{sp_0} \fourier(\T) 
 +\beta \fourier(f_{sp_0})\fourier(\Pleak \T)\\&
 +\alpha g_{sp_0} \fourier(\Pleak)\fourier(\T)
 \end{split}
 \end{equation}

\subsubsection{Separating the random and deterministic terms}	
The modified Green's function considering variability (solution of Equation~\ref{eqn:expanded1}) is not shift-invariant because of process variation (terms $III$ and $IV$), which means that we will not be able to directly convolve the modified Green's function with a power profile without a loss of accuracy.
To overcome this limitation, we split the modified Green's function  into two components: a deterministic component $f_{leaksp}^{det}$, which is shift-invariant and is obtained by assuming the variability to be zero (replacing the baseline leakage power profile with its mean value), and a random component $f_{leaksp}^{rand}$, which is shift-variant and accounts for all the variation in leakage power.

To arrive at the respective expressions for the modified Green's functions, we first split the variable leakage power, $\Pleak$, into two components -- a constant equal to its mean ($\mu$) and a randomly varying part $P_{leak_0}^{var}$. Thus, $\Pleak = \shadedboxb{\mu + P_{leak_0}^{var}}$. Using this in Equation~\ref{eqn:expanded1}, we get:
\begin{equation}
\begin{split}
\label{eqn:expanded2}
 \fourier(\T) &= \fourier(f_{sp_0}) {+} \alpha g_{sp_0} \fourier(\T) 
 {+}\beta \fourier(f_{sp_0})\fourier\bigl(\shadedboxb{(\mu + P_{leak_0}^{var}}) \T\bigr)\\&
 +\alpha g_{sp_0} \fourier\left(\shadedboxb{\mu + P_{leak_0}^{var}}\right)\fourier(\T)\\
 &= \fourier(f_{sp_0}) + \alpha g_{sp_0} \fourier(\T)
 +\mu \beta \fourier(f_{sp_0})\fourier(\T) \\&
 +\beta \fourier(f_{sp_0})\fourier(P_{leak_0}^{var} \T) 
  +\alpha g_{sp_0} \fourier(\mu)\fourier(\T)\\&
 +\alpha g_{sp_0} \fourier(P_{leak_0}^{var})\fourier(\T)\\
 \end{split}
 \end{equation}

 Now, the temperature profile  $\T$ is itself composed of a deterministic and a variable part, $\T = \T^{det} + \T^{var}$.
 The deterministic part can be obtained by assuming the variability to be zero, $P_{leak_0}^{var} = 0$.
 Applying this to Equation~\ref{eqn:expanded2}, we arrive at Equation~\ref{eqn:expanded3}:
  \begin{equation}
\begin{split}
\label{eqn:expanded3}
 \fourier(\T^{det}) &=\fourier(f_{sp_0}) + \alpha g_{sp_0} \fourier(\T^{det})
 +\mu \beta \fourier(f_{sp_0})\fourier(\T^{det})+ \\&
  \alpha g_{sp_0} \fourier(\mu)\fourier(\T^{det})\\
 \end{split}
 \end{equation}

  \begin{equation}
\begin{split}
\label{eqn:detsol}
& \fourier(\T^{det})  =\fourier(f_{leaksp}^{det}) = 
 \frac{\fourier(f_{sp_0})}{1-\alpha g_{sp_0}(1+\fourier(\mu)) -\mu \beta \fourier(f_{sp_0})}
 \end{split}
 \end{equation}
 
 Next, we use an equation similar to Equation~\ref{eqn:leakvar2} for $\fourier(P_{leak_0}^{var} \T)$. We get:
 
   \begin{equation}
\begin{split}
\label{eqn:randTsplit}
  \fourier(P_{leak_0}^{var} \T) &= \fourier(\T)\bigl(\underbrace{P_{leak_0}^{var} -  P_{leak_0}^{var}(\infty,\infty) + P_{leak_0}^{var}(0,0) }_{=Q_{leak_0}^{rand}}\bigr)\\
  &= \fourier(\T) Q_{leak_0}^{rand}
  \end{split}
 \end{equation} 
 
 Substituting Equation~\ref{eqn:randTsplit} and  $\T = \T^{det} + \T^{var}$ in Equation~\ref{eqn:expanded2} and simplifying, we finally get:
 

 \begin{equation}
\begin{split}
\label{eqn:expanded5}
 \fourier(\T^{det})&\left(\shadedboxc{1-  \alpha\left(1+\fourier(\mu)\right)g_{sp_0}
 -\mu \beta \fourier(f_{sp_0})} - \right.\\&
 \left.\beta \fourier(f_{sp_0})Q_{leak_0}^{rand} 
-\alpha \fourier(P_{leak_0}^{var})g_{sp_0} \right) +  \\
 \fourier(\T^{var})&\left(1-  \alpha(1+\fourier(\mu))g_{sp_0}
 -\mu \beta \fourier(f_{sp_0}) \right.\\&- 
 \beta \fourier(f_{sp_0})Q_{leak_0}^{rand} 
-\left.\alpha \fourier(P_{leak_0}^{var})g_{sp_0}\right)  \\ & = \shadedboxc{\fourier(f_{sp_0})}
 \end{split}
 \end{equation}

Now, we substitute the expression for $\fourier(f_{sp_0})$ from Equation~\ref{eqn:expanded3} in Equation~\ref{eqn:expanded5}. After cancelling the common terms (shaded/green ones) we get: 
 \begin{equation}
\begin{split}
\label{eqn:expanded6}
 \fourier(\T^{var})&\left(1{-}  \alpha(1{+}\fourier(\mu))g_{sp_0}
 {-}\mu \beta \fourier(f_{sp_0}) \right.{-} 
 \beta \fourier(f_{sp_0})Q_{leak_0}^{rand} \\&
-\left.\alpha \fourier(P_{leak_0}^{var})g_{sp_0}\right) \\ & =  \fourier(\T^{det})\left( 
 \beta \fourier(f_{sp_0})Q_{leak_0}^{rand}+\alpha \fourier(P_{leak_0}^{var})g_{sp_0}\right) \\
 \end{split}
 \end{equation}
 
  \begin{equation}
\begin{split}
\label{eqn:randsol}
 &\fourier(\T^{var}) = \fourier(f_{leaksp}^{rand})\\ & = \fourier(\T^{det})\times   \frac{\left(  \beta \fourier(f_{sp_0})Q_{leak_0}^{rand} + \alpha \fourier(P_{leak_0}^{var})g_{sp_0}\right) }{\splitfrac{\Bigl(1-  \alpha\Bigl(1+\fourier(\mu)+\fourier(P_{leak_0}^{var})\Bigr)g_{sp_0}
  \Bigr.}{\Bigl.-\mu \beta \fourier(f_{sp_0}) -
 \beta \fourier(f_{sp_0})Q_{leak_0}^{rand} \Bigr)}}
 \end{split}
 \end{equation}
 
The deterministic part of the Green's function remains the same for every variation profile. Hence, it needs to be computed once for a given chip only. The random part needs to be recomputed every time we get a new variation map.

\subsubsection{Full-chip steady-state thermal profile}
The total temperature profile is a sum of the random and deterministic components. The standard approach to compute the deterministic thermal profile is to convolve the deterministic Green's function (Equation~\ref{eqn:detsol}) with the respective power profile. 
The random component of the thermal profile is not shift-invariant and cannot be computed using the convolution operation, since it is location-dependent. Hence, we compute the Hadamard product of the random component with the leakage power profile and scale it by the total dynamic power applied to the chip. This is an approximation that we justify empirically after conducting exhaustive experiments.

Thus the total thermal profile is given by:
   \begin{equation}
\begin{split}
\label{eqn:detrandsol}
 \T = f_{leaksp}^{det} \star P_{dyn} + f_{leaksp}^{rand}* P_{leak_0}^{var}* \sum_{i=1}^{n}\sum_{j=1}^{n} P_{dyn(i,j)} 
 \end{split}
 \end{equation}
where $*$ represents the Hadamard product, $\star$ represents the convolution operation, and $n$ represents the number of grid points in the chip in one direction.

\subsection{Modified Green's function for the Transient Case}
\label{sec:tran}
Next, we look at the temporal evolution of temperature.
Because of the complexity involved in obtaining the transient solution, we do not split the transient Green's function into shift-variant and shift-invariant components. 

The basic transient Green's function equation is given by \cite{lightsim}:
\begin{equation}
\label{eqn:Ttran}
\T = f_{sp} \star P - C f_{sp}\star \frac{\partial \T}{\partial t},
\end{equation}
where $C$ is the thermal capacitance. 

Proceeding in the same manner as the steady-state solution, and looking at Equation~\ref{eqn:leakvar2}, we arrive at the following equation for the transient case:

\begin{equation}
\begin{split}
\label{eqn:FTftran}
\fourier&(\T)  =\fourier(f_{sp_0}) + \alpha g_{sp_0} \fourier(\T) 
 +\beta \fourier(f_{sp_0})\fourier(\Pleak \T)+\\&
 \alpha g_{sp_0} \fourier(\Pleak)\fourier(\T)
 {-} C \bigl(\fourier(f_{sp_0}) {+} 
\alpha \underbrace{\fourier(f_{sp_0}\T)}_{G(u,v)}\bigr) \fourier \left(\frac{\partial \T}{\partial t}\right)\\
&= \fourier(f_{sp_0}) + \alpha g_{sp_0} \fourier(\T) 
 +\beta \fourier(f_{sp_0})\fourier(\Pleak \T)+\\&
 \alpha g_{sp_0} \fourier(\Pleak)\fourier(\T) {-} C \bigg(\fourier(f_{sp_0}) {+}
\alpha g_{sp_0}\fourier(\T) \bigg) \fourier \left(\frac{\partial \T}{\partial t}\right)\\
\end{split}
\end{equation}

The first three terms on the RHS correspond to the steady-state temperature profile, $\T_{ss}$. 
Thus we have:
\begin{equation}
\begin{split}
\fourier(\T) =& \fourier(\T_{ss})- C \bigl(\fourier(f_{sp_0}) + 
\shadedbox{\alpha \fourier(\T)g_{sp_0} } \bigr) \fourier \left(\frac{\partial \T}{\partial t}\right)\\
=& \fourier(\T_{ss})- C \fourier(f_{sp_0})\left(\frac{\partial \fourier(\T)}{\partial t}\right) \\& -\shadedbox{\alpha C 
 \fourier(\T)g_{sp_0}    \left(\frac{\partial \fourier(\T)}{\partial t}\right)}\\
\end{split}
\end{equation}
The shaded term is of the form $\fourier(\T)\frac{\partial \fourier(\T)}{\partial t}$, making the solution complex.

We separate the partial derivative term next:
\begin{equation}
\begin{split}
 \frac{\partial \fourier(\T)}{\partial t}& = -\frac{\fourier(\T) - \fourier(\T_{ss})}{C\fourier(f_{sp_0}) + 
\alpha C \fourier(\T)g_{sp_0}  }
\end{split}
\end{equation}

Separating the variables and replacing partial derivatives with total derivatives (since there is only one variable):
\begin{equation}
\begin{split}
 \frac{C\fourier(f_{sp_0}) + 
\alpha C \fourier(\T)g_{sp_0}}{\fourier(\T) - \fourier(\T_{ss})}d \fourier(\T)& = -d t
\end{split}
\end{equation}
Integrating on both sides:
\begin{equation}
\label{eqn:traniter}
\begin{split}
&C\fourier(f_{sp_0}) ln(\fourier(\T) - \fourier(\T_{ss})) + 
\alpha C g_{sp_0}(\fourier(\T) - \fourier(\T_{ss})) +\\& \alpha C g_{sp_0}\fourier(\T_{ss})ln(\fourier(\T) {-} \fourier(\T_{ss}))  = -(t {+} C_1)\\
\\
&ln(\fourier(\T) - \fourier(\T_{ss}))\bigl(C\fourier(f_{sp_0}) + \alpha C g_{sp_0}\fourier(\T_{ss})\bigr)\\& + 
\alpha C g_{sp_0}(\fourier(\T) - \fourier(\T_{ss})) = -(t + C_1)\\
\end{split}
\end{equation}
We need to solve for $\fourier(\T)$ here.
We observe that the last term on the LHS is small since both $\alpha$ and $C$ are small numbers. Hence, we ignore this term.

Thus we have:
\begin{equation}
\label{eqn:tranlneqn}
\begin{split}
ln(\fourier(\T) {-} \fourier(\T_{ss}))\bigl(C\fourier(f_{sp_0}) {+} \alpha C g_{sp_0}\fourier(\T_{ss})\bigr) = -(t {+} C_1)\\
\end{split}
\end{equation}

Taking the exponential on both sides and simplifying:
\begin{equation}
\label{eqn:tranexpeqn}
\begin{split}
\fourier(\T) &= \fourier(\T_{ss}) + e^{-\frac{t+C_1}{C\fourier(f_{sp_0}) + \alpha C g_{sp_0}\fourier(\T_{ss})}} \\
&=\fourier(\T_{ss}) + k_1 e^{-\frac{t}{C\fourier(f_{sp_0}) + \alpha C g_{sp_0}\fourier(\T_{ss})}}
\end{split}
\end{equation}
where $k_1$ is a constant.

To compute $k_1$, we see that at $t=0$, the temperature rise $\T$ is zero. Substituting these in Equation~\ref{eqn:tranexpeqn}, we get $k_1 = -\fourier(\T_{ss})$.

Thus the final transient leakage and variation-aware Green's function is given by:
\begin{equation}
\begin{large}
\label{eqn:tranGreen}
\boxed{\fourier(\T) {=} f_{leaksp}^{tran} {=} \fourier(\T_{ss}) \bigl(1 - e^{-\frac{t}{C\fourier(f_{sp_0}) + \alpha C g_{sp_0}\fourier(\T_{ss})}}\bigr) }\\
\end{large}
\end{equation}

\subsection{Full-chip Transient Thermal Profile for Time-varying Sources}
\label{sec:truetran}
Next, we use the transient Green's function to compute the full-chip thermal profile corresponding to a time-varying power profile.
Note that the transient Green's function is a 3D function of space as well as time. 

We observe that the thermal response decays to under 1\% of its peak value within $5~ms$ after a power source is removed. Thus, we conclude that incorporating the thermal response corresponding to the power sources in the last $5~ms$ only should be sufficient to attain a reasonable accuracy. In the general case, we need to consider the power sources from the last $k$ time instants.

Let $P(t_i)$ denote the
instantaneous power profile at time $t_i$, and let $\T(t_i)$ be the temperature profile. 
The Green's function is obtained by applying a power source of unit magnitude and $1~ms$ width at $t=0$ at the center of the chip. The resulting thermal response is measured for the entire chip at intervals of 1~ms, resulting in a 3D tensor. 

We start with an initial temperature $T_0$, and at every time instant, we either increase or decrease the temperature depending on the change in power values in each time interval. To determine the amount of change in temperature, we compute a difference of the power profiles at time instant $t_i$ and $t_{i-1}$ and convolve this with the corresponding leakage-aware Green's function sampled at $t_i$. We do this for each of the last $k$ time steps, and sum the effects up. Beyond $k$ time steps, the power values in the past would have achieved steady-state and their effects would already have been considered.
Let us first describe our full-chip transient estimation approach mathematically for a 2D chip without leakage. Without any loss of generality, let us assume the initial temperature $T_0$ to be zero.

Let the Green's function at time instant $t_0$ be denoted by $f_{sp}(x,y,t_0)$. Let the power dissipation profile be $P(x,y,t)$. 

Let us apply a power source at $t=0$.

At $t=0$, the temperature rise is given by:
\begin{equation}
\label{eqn:tranapprox}
\begin{split}
\T(t_1) &= {fsp(t_1)} \star \left({P(t_1) - P(t_0)}\right)\\
&= {fsp(t_1)} \star {(P(t_1) - 0)}\\
\T(t_2) &= {fsp(t_1)} \star \left({P(t_2) - P(t_1)}\right) + {fsp(t_2)} \star \left({P(t_1) -  0}\right)\\
\T(t_3) &= {fsp(t_1)} \star \left({P(t_3) {-} P(t_2)}\right) {+}
 {fsp(t_2)} {\star} \left({P(t_2) {-} P(t_1)}\right)\\& + {fsp(t_3)} \star \left({P(t_1) -  0}\right)\\
...\\
\T(t_n) &= {fsp(t_1)} \star ({P(t_n) {-} P(t_{n-1})}) + {fsp(t_2)} \star \left(P(t_{n-1}) {-}\right. 
\\& \left. P(t_{n-2})\right) + ... +{fsp(t_{5})} \star({P(t_{n-5}) {-} P(t_{n-6})}) + \\& {fsp(t_{6})} \star({P(t_{n-6}) {-} P(t_{n-7}}) + ... + {fsp(t_{n})} \star \\& {(P(t_1) {-} P(t_0))}\\
\\
\end{split}
\end{equation}

Now, beyond $5~ms$, we assume that the step response saturates. 

Thus 
$fsp(t_{5}) = fsp(t_{6}) = fsp(t_{7}) .... =fsp(t_{\infty})$. 

Equation~\ref{eqn:tranapprox} reduces to:
\begin{footnotesize}
\begin{equation}
\begin{split}
\T(t_n) &\approx {fsp(t_1)} \star ({P(t_n) {-} P(t_{n-1})}) {+} {fsp(t_2)} \star \\
&\left({P(t_{n-1}) {-} P(t_{n-2})}\right)  + ... +{fsp(t_{\infty})} \star \\ & \bigl(\underbrace{{(P(t_{n-5}) {-} P(t_{n-6}) {+} P(t_{n-6}) {-} P(t_{n-7})  {+ ... +} 
P(t_1) {-} P(t_0))}}_{=P(t_{n-5})}\bigr)
\end{split}
\end{equation}
\end{footnotesize}

We can see that all terms cancel each other in the third convolution term, and only $P(t_{n-5})$ remains. \\
Thus we can calculate
the transient thermal profile by considering the last 5 time instants only.
To incorporate leakage, we simply need to substitute the \emph{modified leakage aware} Green's functions in place of the basic Green's function.

\subsection{Thermal Estimation at the Edges and Corners}
In the Green's function approach, the edges and corners have to be handled separately. The standard procedure to do so is to use an analogy with the method of images from 
electromagnetics (also used in \cite{powerblur2014}). In this approach, the power matrix is extended to twice its size and padded with mirror image sources on the other side of the boundary at an equal distance from the edge.
To compute the full-chip thermal profile, we convolve the modified deterministic Green's functions with the 
dynamic power profile and multiply the random Green's function with the baseline leakage power profile (Equation~\ref{eqn:detrandsol}).

%% file: evaluation.tex
\section{Evaluation}
\label{sec:eval}

\subsection{Setup}
We use an augmented version of the thermal modeling tool Hotspot~\cite{hotspot6} to carry out 
simulations with variable leakage power and conductivity. The scripts to invoke HotSpot has been written in R. 
We run all our HotSpot 
simulations on an Intel i7-7700 4-core CPU running Ubuntu 16.04 with 16 GB of RAM. 
We implemented and tested our proposed algorithm in Matlab on a Windows 8 desktop 
with an Intel i7-2600S processor and 8 GB of RAM.
We discretized the chip into a $64\times64$ grid. The parameters of the modeled chip are given in Table~\ref{tab:param}.

\begin{table}[!htb]
\small
 \begin{center}
 \caption{Parameters of the chip~\cite{hotspot6}\label{tab:param}}
 \begin{tabular}{p{5cm}l}
\toprule
Parameter & Value \\
\midrule
No. of grid points per layer, $n$		 & $64\times64$ \\
$\beta$ 	 & 0.0275  \\
Die size & 100 mm$^2$ \\
Die thickness & 0.15 mm \\
Nominal Silicon conductivity & 130 W/m-K \\
TIM thickness & 0.02 mm \\
TIM conductivity & 4 W/m-K \\
Spreader thickness & 3.5 mm\\
Spreader conductivity & 400 W/m-K \\
Ambient temperature & 318.15K \\
\bottomrule
\end{tabular}
\end{center}
\end{table}

\subsubsection*{Error Metric}
We use the mean absolute error and the percentage error relative to the maximum temperature \textbf{rise} (calculated temperature minus the ambient temperature) as the error metric. Other thermal modeling tools often report errors relative to the absolute maximum temperature in the die, which under-represents the error~\cite{powerblur2014, isac}.

\subsection{Calibration of the Setup}
To calibrate our HotSpot setup, we use the commercial CFD software Ansys Icepak. It is an industry-standard 
tool widely used for high-accuracy thermal simulations. We model an identical layout in HotSpot and Icepak 
and compare the temperature values obtained from the two tools. We find that the normalized temperature 
values obtained using both of these tools conform well (within 1.5\%).

\subsection{Steady-State Results}

\begin{table}[!htbp]
  \begin{tabular}{p{4.3cm}|p{1.5cm}|p{1.9cm}}
    \toprule
    {Simulator} &
      {Considering $\kappa(T)$} &
      {Without considering $\kappa(T)$} \\
      \midrule
    Hotspot\footnotemark & $18~minutes$ & $4~s$   \\
    3D-ICE & -- & $1.36~s$   \\
    Icepak & $15~minutes$ & $15~minutes$  \\
    Jaffari et al.\cite{jaffari} & -- & $158~s$ \\
    \rowcolor{lblue}
    \textbf{\fname}& \textbf{$2.9~ms$} & \textbf{$2.9~ms$}  \\	  
    \rowcolor{vlblue}
    {~~\fname det. Green's func. {(Offline)}}& \textbf{$0.55~ms$} & \textbf{$0.55~ms$}  \\	  
    \rowcolor{vlblue}
    {~~\fname rand. Green's func. {(Offline)}}& \textbf{$1.6~ms$} & \textbf{$1.6~ms$}  \\	  
    \rowcolor{vlblue}
        {~~\fname full-chip (Online)}& \textbf{$0.74~ms$} & \textbf{$0.74~ms$}  \\	  

    \bottomrule
  \end{tabular}
  \footnotesize{\\1. To model temperature-dependent conductivity, detailed thermal modeling is done in HotSpot, since the properties of each block are different. \\2. HotSpot, 3D-ICE and Icepak do not consider variability}\\
  \caption{Speed of the studied simulators  \label{tab:exectime}}	
  	\end{table}

\begin{table*}[!htbp]
\centering
  \caption{Errors in various scenarios}	
  \label{tab:error}
  \setlength\tabcolsep{3pt} 
  \begin{tabular}{p{4.1cm}|ccc|ccc}
    \toprule
    \multirow{2}{*}{Effects considered} &
      \multicolumn{3}{c|}{Test Case 1 (Alpha21264)} &
      \multicolumn{3}{c}{Test Case 2 } \\
      & {Max. Temp. (K)} & {Max. Deviation (K)} & {Percent Deviation} & {Max. Temp. (K)} & {Max. Deviation (K)} & {Percent Deviation} \\
      \midrule
	\cmnt{5}No effects&				341.36	& 	\br{6.77} 	&	22.6	&372.90 & 9.01	&14.1\\
	\cmnt{3}Rand.-cond., Cond.(T)  &	341.86	&	\br{6.27}	&	20.9	&377.59&4.32& 6.8\\
	\cmnt{8}Leakage-var&					344.04	&	\dblue{4.09}	&	13.6	&375.29&6.62& 11.6\\
	\cmnt{7}Leakage(T)&			344.30 	& 	\dblue{3.83} 	&	12.8 	&374.56& 7.35&12.2\\
	\cmnt{6}Rand.-cond., Cond.(T), Leakage(T)&	344.91& \dblue{3.22}&10.7& 378.39 &3.52 &5.8\\
	\cmnt{2}Leakage-var, Leakage(T)&	347.43	&	0.70	&	2.33		&377.16&4.75&7.5\\
	\cmnt{4}Cond.(T), Leakage-var, Leakage(T)&	348.12&	0.01&	0.03&381.36 & 0.55&0.86\\
	\cmnt{1}Rand.-cond., Cond.(T), Leakage-var, Leakage(T)&	348.13& -- & -- &381.91&--&--\\
	\hline
	\fname 			&	\textbf{347.58}&	\textbf{0.55} &\textbf{1.8}& \textbf{379.31}& \textbf{2.60}&\textbf{4.1}\\
    \bottomrule
  \end{tabular}	
  \footnotesize{\\Leakage(T) = temperature-dep. leakage, Leakage-var = variability in leakage, Rand.-cond. = random conductivity, cond.(T) = temperature-dep. conductivity}\\
\end{table*}

\begin{figure*}[!htb]
	\begin{subfigure}[b]{0.7\columnwidth}
		\begin{center}
		\includegraphics[width=0.85\columnwidth, trim = {1cm 6cm 2cm 6cm},
		clip=true]{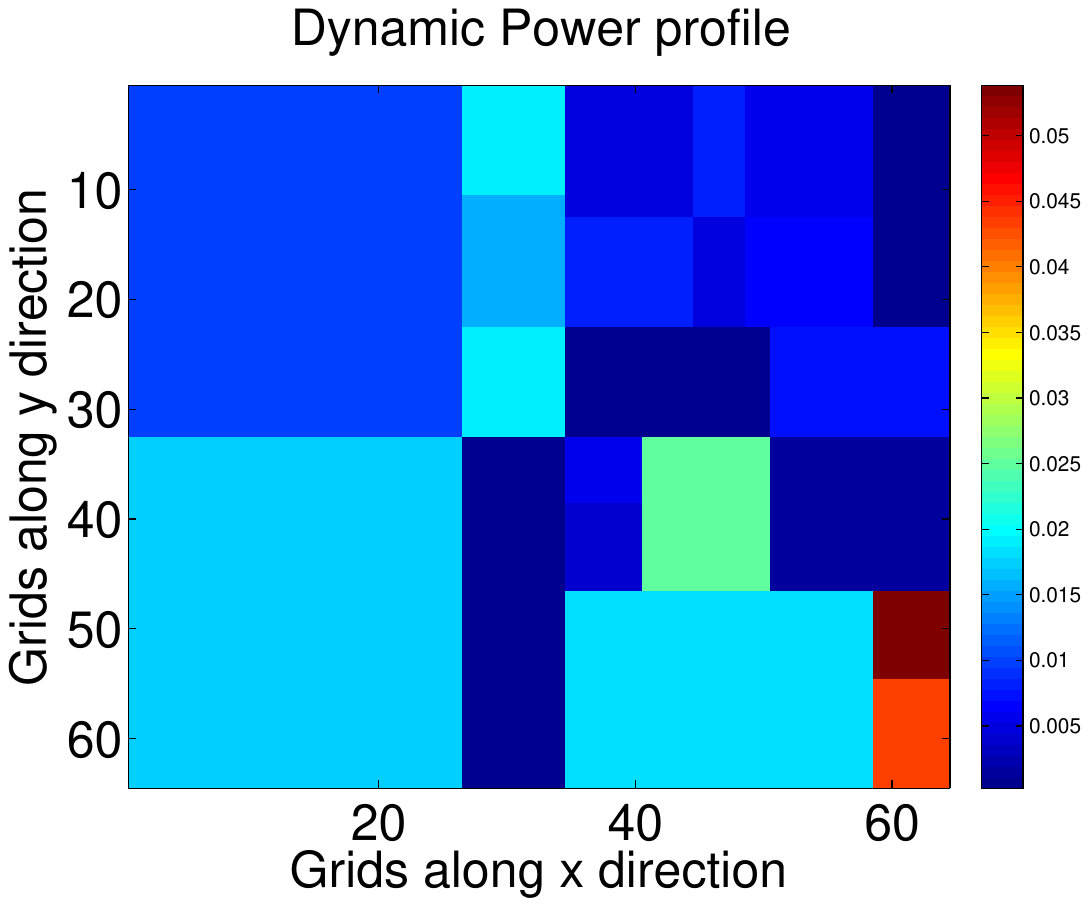}
		\end{center}
		\subcaption{Dynamic Power} \label{fig:dynpower}
	\end{subfigure}%
	\captionsetup{subrefformat=parens}
        \begin{subfigure}[b]{0.7\columnwidth}
		\begin{center}
                \includegraphics[width=0.9\columnwidth, trim = {1.cm 6cm 0.4cm 6.5cm},
                clip=true]{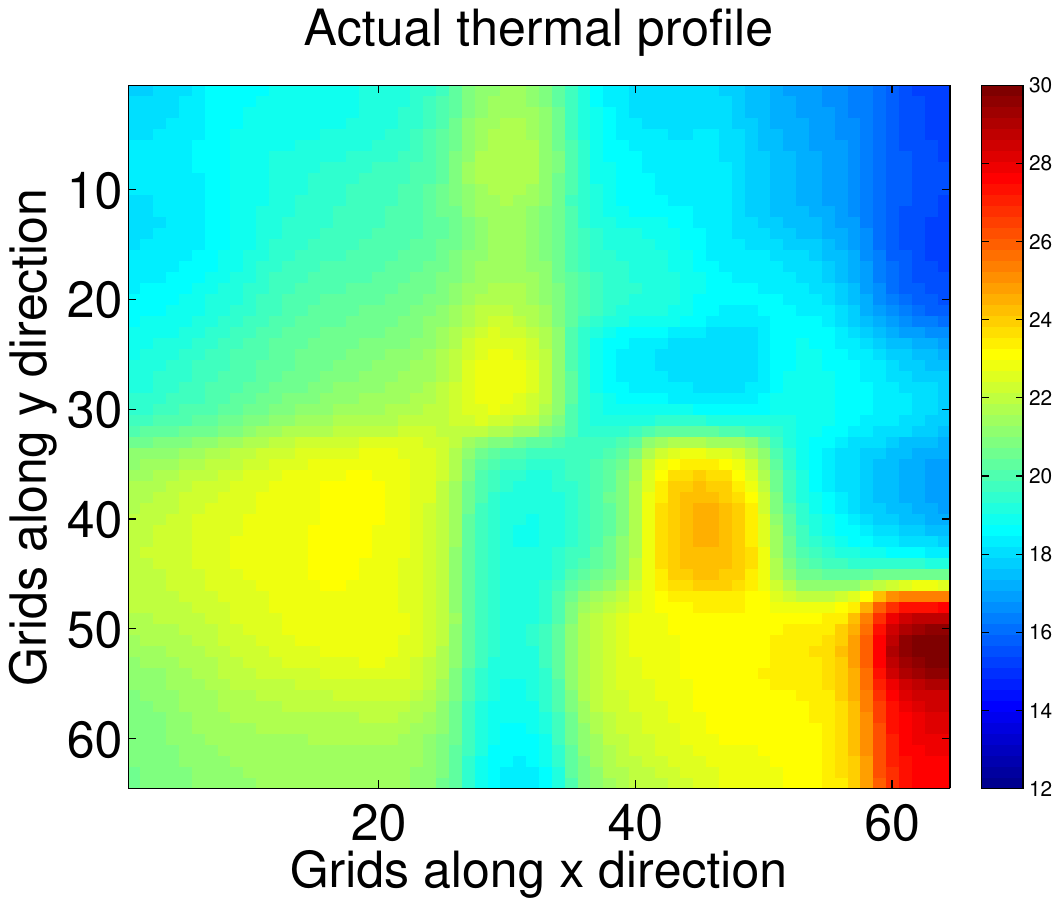}
		\end{center}
                \subcaption{Calculated thermal profile} \label{fig:calctemp}
        \end{subfigure}%
                \begin{subfigure}[b]{0.7\columnwidth}
		\begin{center}
                \includegraphics[width=0.9\columnwidth, trim = {.8cm 6cm 0.4cm 6.5cm},
                clip=true]{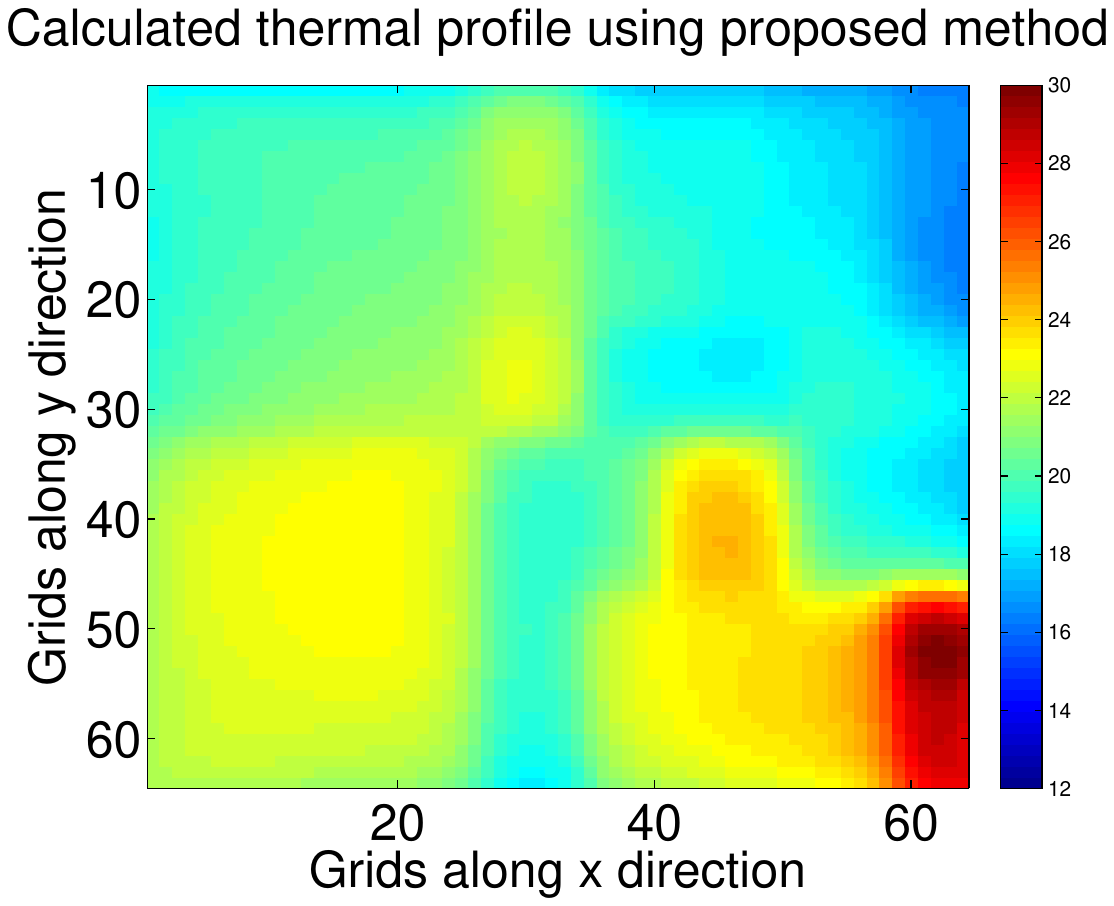}
		\end{center}
                \subcaption{Actual Thermal profile} \label{fig:acttemp}
        \end{subfigure}
        \captionsetup{subrefformat=parens}
        \caption{Evaluation for test case 1: Alpha21264}\label{fig:results}
\end{figure*}

\begin{figure*}[!htb]
	\begin{subfigure}[b]{0.7\columnwidth}
		\begin{center}
		\includegraphics[width=0.85\columnwidth, trim = {1cm 6cm 2cm 6cm},
		clip=true]{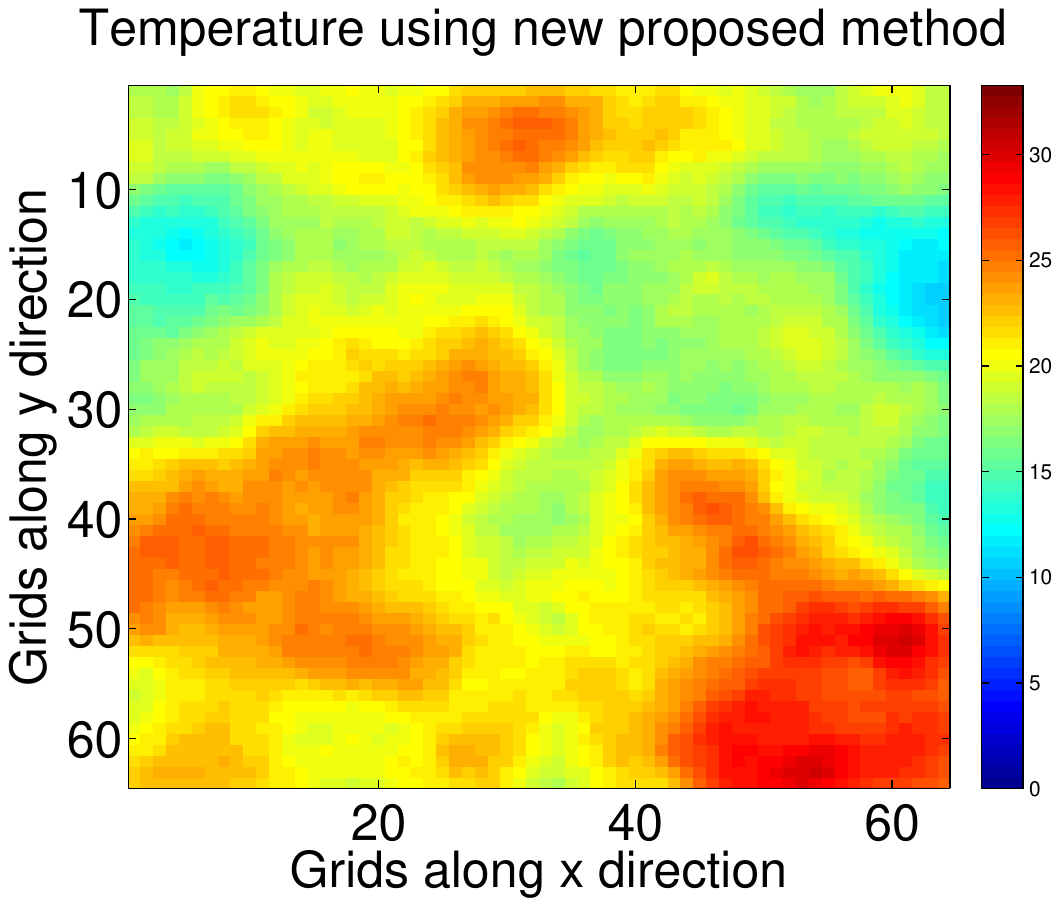}
		\end{center}
		\subcaption{Calculated thermal profile} \label{fig:calctc3}
	\end{subfigure}%
	\captionsetup{subrefformat=parens}
        \begin{subfigure}[b]{0.7\columnwidth}
		\begin{center}
                \includegraphics[width=0.9\columnwidth, trim = {1.cm 6cm 0.4cm 6.5cm},
                clip=true]{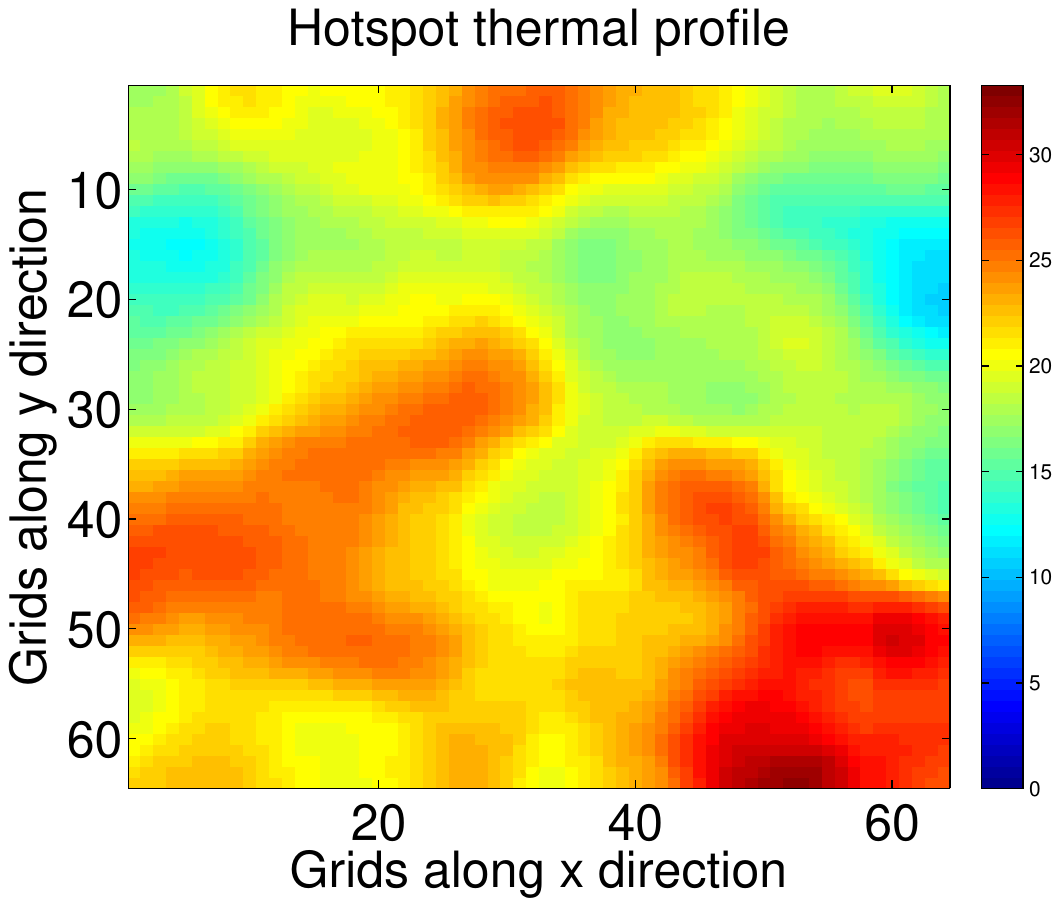}
		\end{center}
                \subcaption{HotSpot thermal profile} \label{fig:acttc3}
        \end{subfigure}%
                \begin{subfigure}[b]{0.7\columnwidth}
		\begin{center}
                \includegraphics[width=0.9\columnwidth, trim = {.8cm 6cm 0.4cm 6.5cm},
                clip=true]{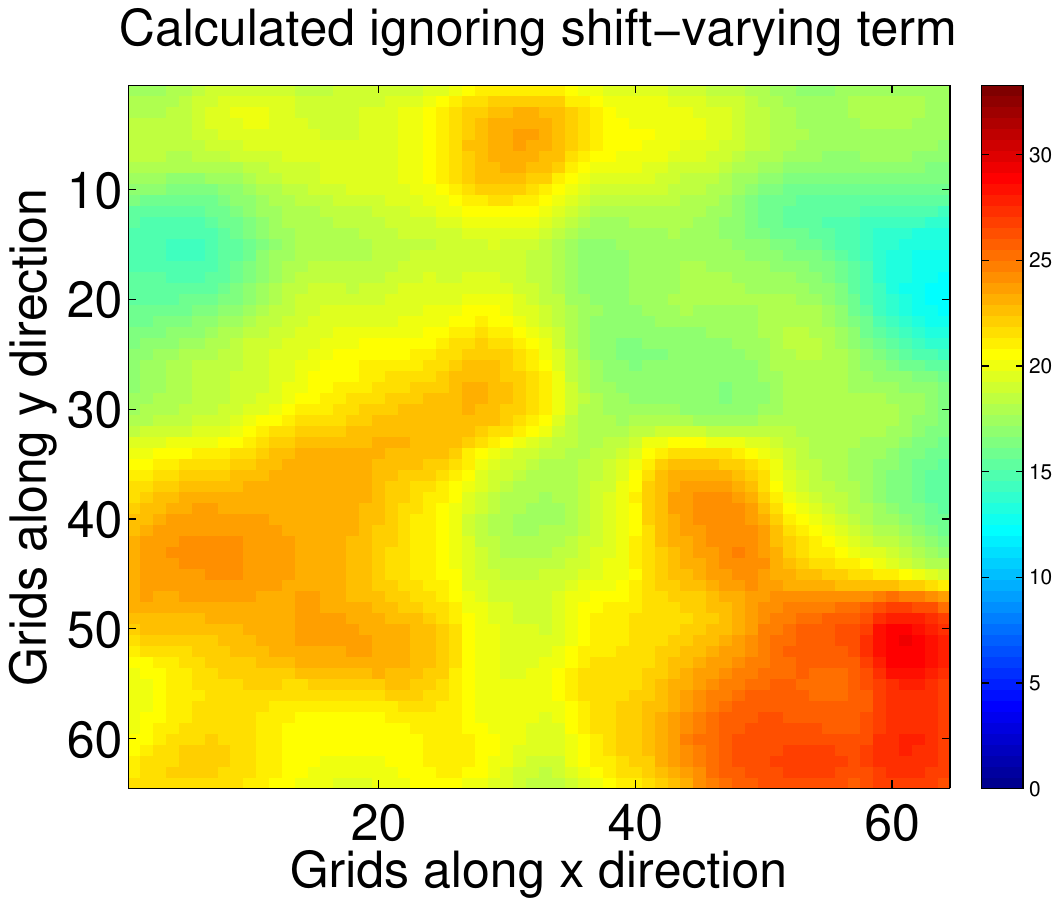}
		\end{center}
                \subcaption{Calculated thermal profile without $f_{leaksp}^{rand}$} \label{fig:igntc3}
        \end{subfigure}
        \captionsetup{subrefformat=parens}
        \caption{Evaluation for test case 3: high variance}\label{fig:results2}
\end{figure*}

\begin{figure}
\centering
\includegraphics[trim={0 0 0 0},clip,width= \columnwidth]{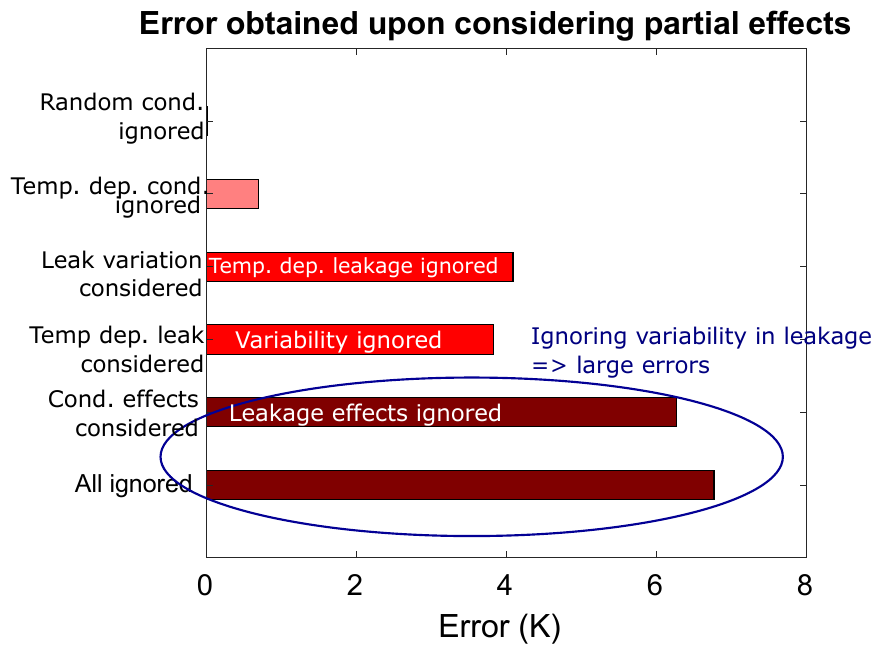}
\caption{Error in various scenarios: considering variability in leakage power is as important as considering the temperature dependence of leakage \label{fig:error}}
\end{figure}

\subsubsection{Modified Green's functions for the steady-state}
The first step in our proposed method is to compute the modified Green's functions that we have derived in our work. 
To do so, we first need the baseline leakage power map considering variation at ambient temperature. To obtain the variation-aware baseline leakage power maps, we use the popular 
variation modeling tool, \textit{Varius}~\cite{varius}. Using a similar approach, we model the randomness in conductivity values because of random dopant fluctuations. 

Next, we obtain the corresponding baseline Green's functions using a modified version of HotSpot. We apply a unit impulse power source to the center of the chip and obtain the baseline Green's 
function considering the variability in conductivity. 
We then use Equation~\ref{eqn:detsol} to obtain the deterministic part of the modified Green's function accounting for the 
effects of temperature-dependent conductivity and leakage power. Our approach takes $0.55~ms$ to compute the deterministic modified Green's function.
Next, we compute the random part of the Green's function using Equation~\ref{eqn:randsol}. This takes a further $1.86~ms$. This part needs to be computed once for a chip.

\subsubsection{Full-chip steady-state thermal simulations}
At runtime, the dynamic power profile is mirrored 
and the full-chip thermal profile is computed using the calculated Green's functions using Equation~\ref{eqn:detrandsol}. This step takes an 
additional $0.74~ms$. Thus the total time taken by our algorithm is $2.89~ms$ (online time = $0.74~ms$ + deterministic Green's function computation = $0.55~ms$ + random Green's function computation = $1.6~ms$). The mean absolute error is limited to 
2\% (as demonstrated by multiple test cases, Table~\ref{tab:steadyres}).

To validate our proposal, we adopt the following approach: the leakage power obtained from Varius is added to the dynamic power profile, and HotSpot is invoked 
iteratively. After each iteration, we update the leakage power and conductivity 
values based on the current temperature. 
We keep iterating until the temperature values converge. HotSpot supports modeling of variable conductivity only when detailed 3D modeling is enabled, since different conductivity values for different blocks result in a change of the parameters of the differential equation from block to block. As a result, HotSpot requires $18~min$  to compute the final temperature. If we do not model variable conductivity, the simulation completes within 4s. Thus, our method provides a $370000\times$ speedup over HotSpot in steady-state thermal simulation.

\noindent \textbf{Test Case 1 [Real floorplan]:} We validate our approach using the floorplan of the 
Alpha21264 processor. The 
power values are taken from the \textit{ev6} test case of HotSpot. The leakage and dynamic power profile and 
the corresponding temperature profiles are shown in Figure~\ref{fig:results}.
We can see in Figure~\ref{fig:calctemp} that the calculated thermal profile matches the actual thermal 
profile very well (mean absolute error = $0.36$\celsius, i.e., within 2\%).

\noindent \textbf{Test Case 2 [Stress testing]:} In this case, multiple dynamic power sources are applied to 
different 
locations on the chip. The total dynamic power is $8~W$. Although the total power applied is 
lower than test case 1, the power density of the sources is much higher, resulting in a higher maximum temperature. In this case, too, the temperature obtained 
using our algorithm 
matches the actual value very well, with a mean absolute error limited to 0.7\celsius (1.1\%).

\noindent \textbf{Test Case 3 [Variance testing]:} In this case, we apply the same dynamic power as test case 1, but the baseline leakage power has a higher variance. The calculated thermal profile is shown in Figure~\ref{fig:calctc3}, while the corresponding thermal profile obtained from HotSpot is shown in Figure~\ref{fig:acttc3}. This test case is a limit study. 
We see that here, too, the calculated and the actual thermal profiles match closely. The mean absolute error is 0.61\celsius for a maximum temperature rise of 32.3\celsius (1.9\%), while the maximum temperature is 77.3\celsius. 
The maximum error at the hotspot location is $2.8$\celsius. 
In Figure~\ref{fig:igntc3}, we show the temperature profile if the random component of the solution is ignored (Equation~\ref{eqn:detrandsol}). We see that the errors are larger in this case, and the main hotspot location is missed. The mean absolute error upon ignoring the random effects is 1.2\celsius (3.7\%), while the maximum error is 5.9\celsius (18.3\%). Thus, we are able to lower the error in modeling process variation-aware thermal profile by up to 52\% using our proposed approach. Moreover, our proposed method helps in capturing the location of the hotspot much more accurately, which would otherwise get missed.

\noindent \textbf{Test Case 4 [Variance + Stress testing]:} In this case, we have a uniform power applied to the entire chip. The total power dissipated is $204.8~W$. In this case, the mean absolute error using our proposed approach is $1.5$\celsius for a maximum temperature rise of $77$\celsius (1.9\%). The mean absolute error upon ignoring the shift-varying Green's function goes up to $3$\celsius (3.9\%).

\noindent \textbf{Test Case 5 [Variance + Stress testing]:} In this case, the alternate grid points have power sources applied to them. The total power dissipated is $51.2~W$. Our proposed method results in a mean absolute error of $0.6$\celsius for a maximum temperature rise of $29.2$\celsius (2\%). 

The steady state test cases are summarized in Table~\ref{tab:steadyres}.
\begin{table}[!htbp]
  \begin{tabular}{l|c|c|c}
    \toprule
    {Test Case} & {Total power ($W$)} & Max. temp. (\celsius) & Mean abs. error (\celsius) \\
      \midrule
    Test case 1 & 48.9 & 74.9 & 0.36\celsius (1.2\%)   \\
    {Test case 2}& 8.0 & 108.8 & 0.70\celsius (1.1\%)   \\	  
    Test case 3 & 48.9 & 77.3 & 0.61\celsius (1.9\%)   \\
    Test case 4 & 204.8 & 122.1 & 1.45\celsius (1.9\%)   \\
    Test case 5 & 51.2 & 74.2 & 0.59\celsius (2\%)  \\    
    \bottomrule
  \end{tabular}
  \caption{Steady-state test cases summary \label{tab:steadyres}}	
  	\end{table}

\subsection{Transient Results}

\subsubsection{Modified transient Green's functions}
\begin{figure}
\centering
\includegraphics[trim={0 5cm 0 5cm},clip,width= 0.99 \columnwidth]{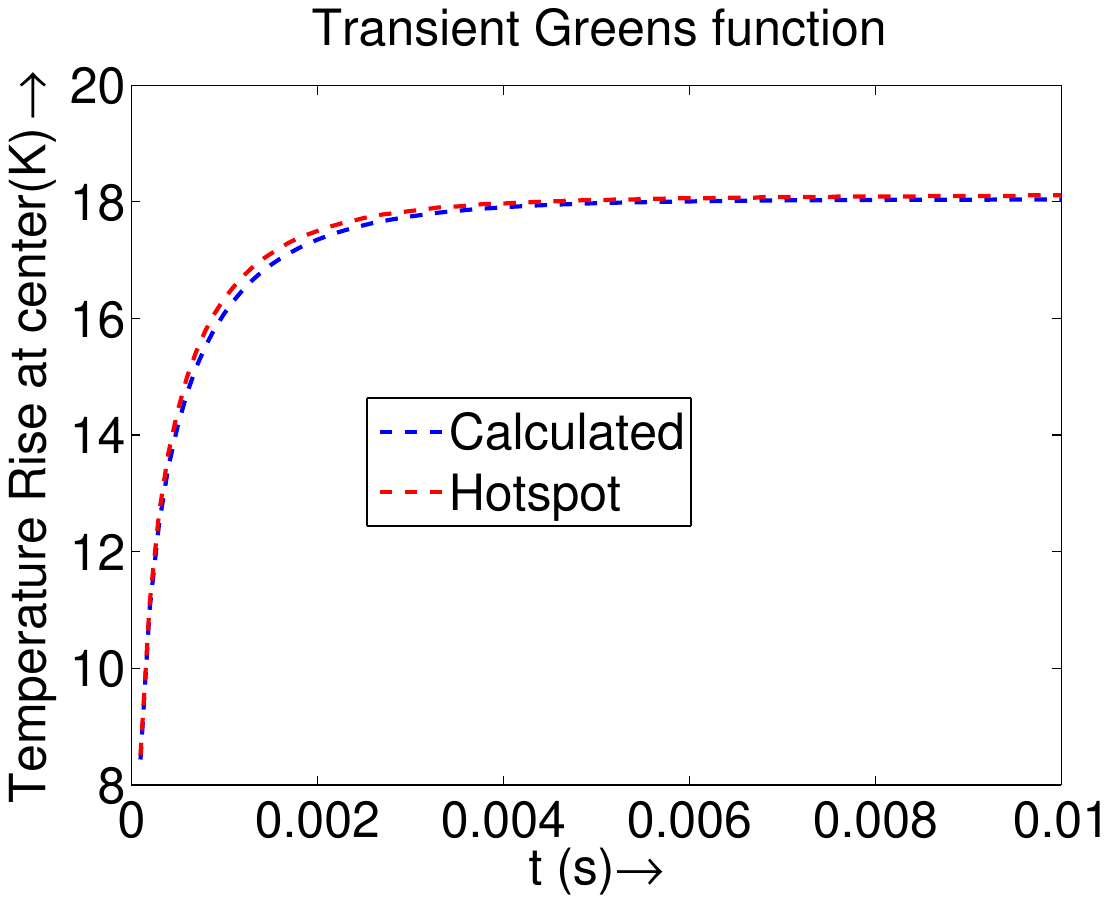}
\caption{Transient Green's function considering all effects \label{fig:Greentrancalc}}
\end{figure}

We obtain the modified transient Green's function using the leakage-aware steady-state Green's function as the starting point. 
Figure~\ref{fig:Greentrancalc} shows the temporal evolution of the calculated transient Green's function using Equation~\ref{eqn:tranGreen} at the center of the chip. The estimation error is less than 3\% at all times. We compute the modified Green's function at 100 time instants between 0 and $10~ms$.
Our algorithm takes $0.12~s$ to compute the temperature profiles for the 100 time steps (Table~\ref{tab:exectimetran}).

\begin{table}[!tbp]
  \begin{tabular}{l|c}
    \toprule
    {Simulator} &
      {Time } \\
      \midrule
    Hotspot & $18-20~minutes$    \\
    \rowcolor{lblue}
    \textbf{\fname}& \textbf{$0.29~s$ (150 time steps)}   \\	  
    \rowcolor{vlblue}
    {~~\fname modified Green's func.}& \textbf{$120~ms$} (100 time steps)  \\	  
    \rowcolor{vlblue}
    {~~\fname full chip step response}& \textbf{$70~ms$} (100 time steps)   \\	  
    \rowcolor{vlblue}
        {~~\fname full chip time varying temp.}& \textbf{$290~ms$} (150 time steps) \\	  

    \bottomrule
  \end{tabular}
  \footnotesize{\\1. To model temperature-dependent conductivity, detailed thermal modeling is done in HotSpot, since the properties of each block are different. \\2. HotSpot, 3D-ICE and Icepak do not consider variability}\\
  \caption{Speed of existing simulators  \label{tab:exectimetran}}	
  	\end{table}

\subsubsection{Full chip transient thermal profile}
\begin{figure*}[!htbp]
\begin{subfigure}[b]{0.24\linewidth}
	\centering
	\includegraphics[width=0.9\linewidth , trim = {.8cm 6.5cm 2cm 6.5cm}, clip=true ]{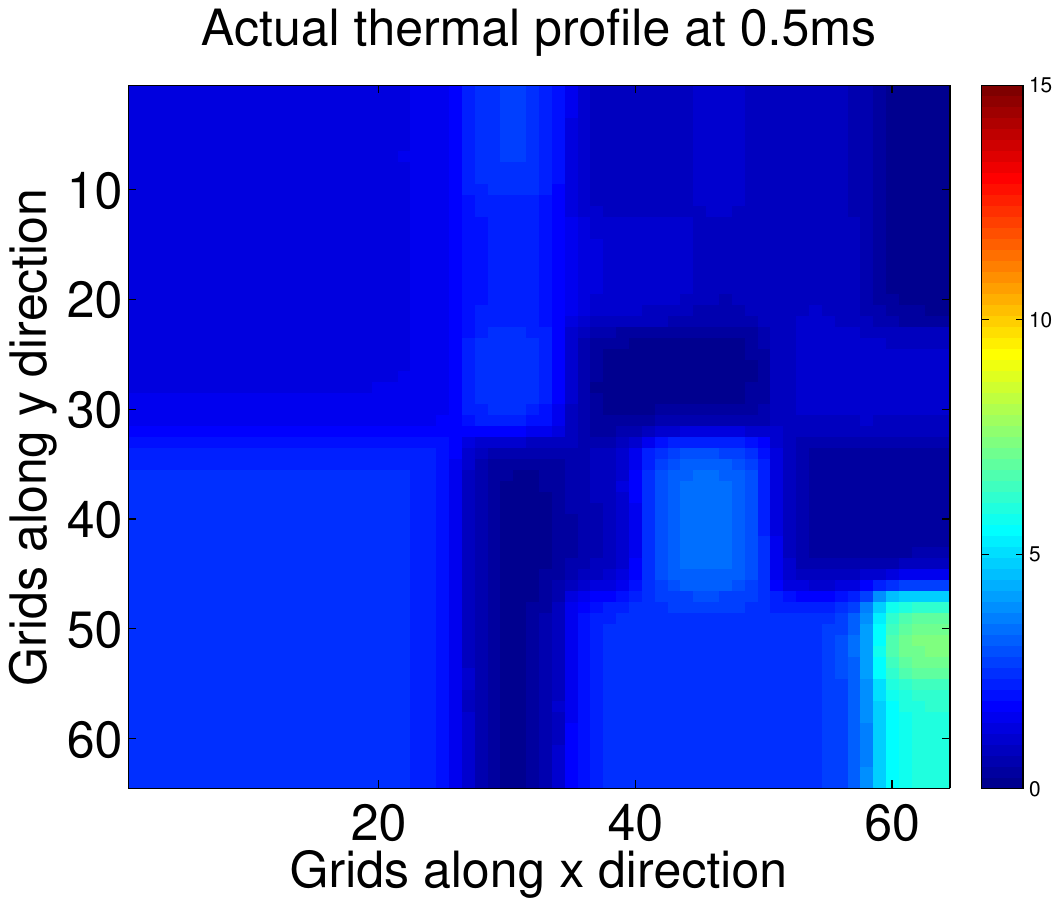}
	\caption{Temperature rise at $0.5~ms$}\label{fig:Ttranfullevol0}
\end{subfigure}
                \begin{subfigure}[b]{0.24\linewidth}
	\centering
	\includegraphics[width=0.9\linewidth , trim = {.8cm 6.5cm 2cm 6.5cm}, clip=true ]{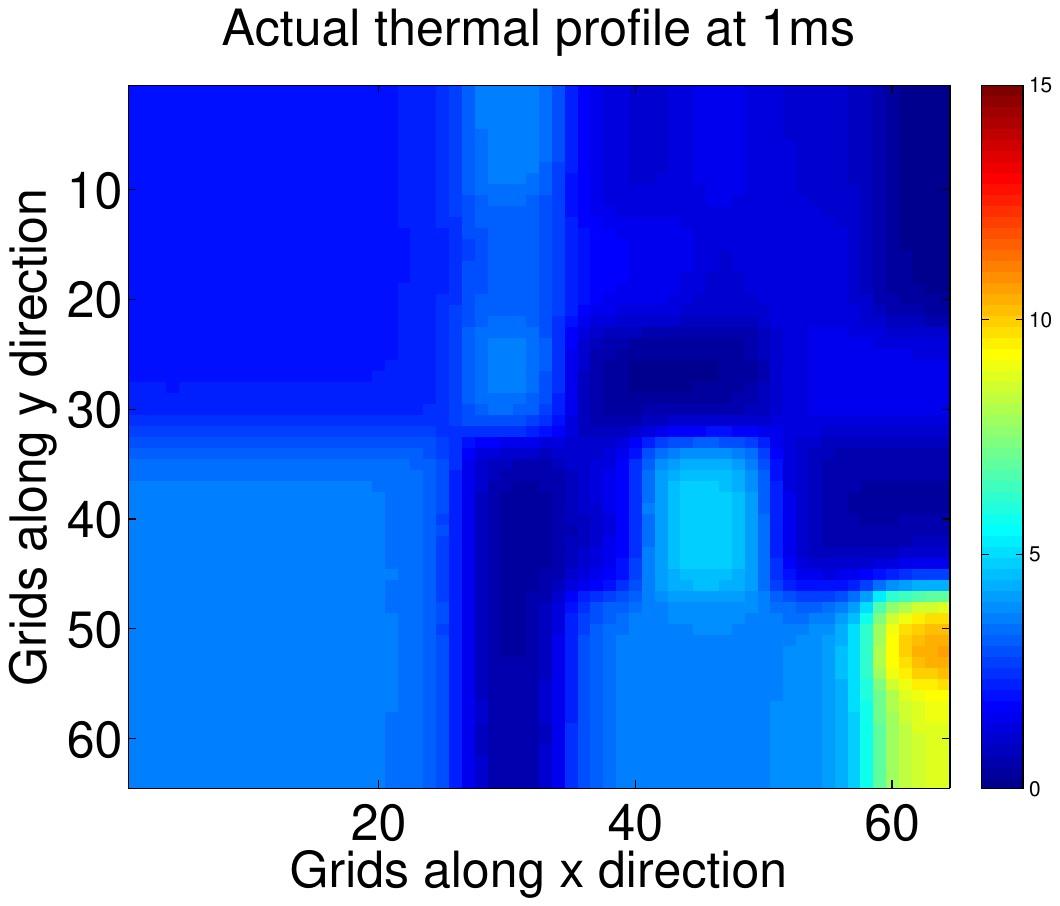}
	\caption{Temperature rise at $1~ms$}\label{fig:Ttranfullevol1}
\end{subfigure}
                \begin{subfigure}[b]{0.24\linewidth}
	\centering
	\includegraphics[width=0.9\linewidth , trim = {.8cm 6.5cm 2.0cm 6.5cm}, clip=true ]{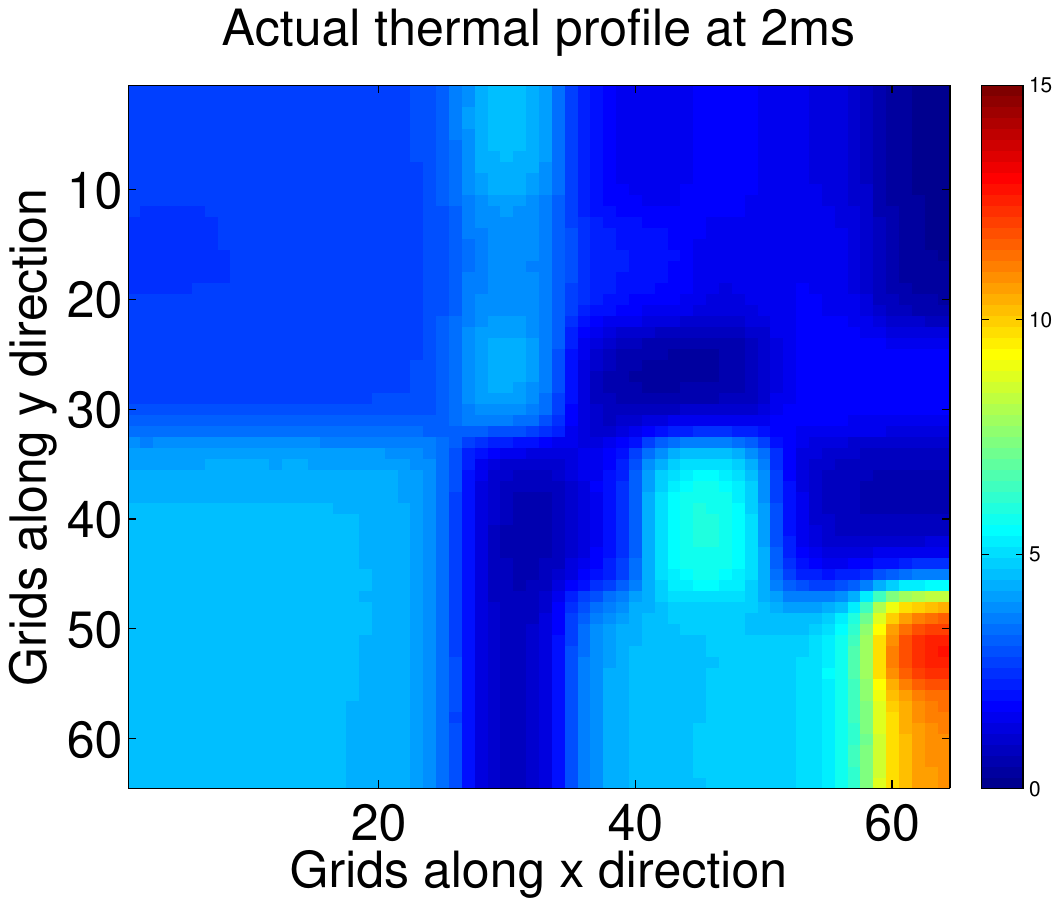}
	\caption{Temperature rise at $2~ms$}\label{fig:Ttranfullevol2}
\end{subfigure}
\begin{subfigure}[b]{0.24 \linewidth}
	\centering
	\includegraphics[width=0.9\linewidth , trim = {.8cm 6.5cm 2.0cm 6.5cm}, clip=true
	]{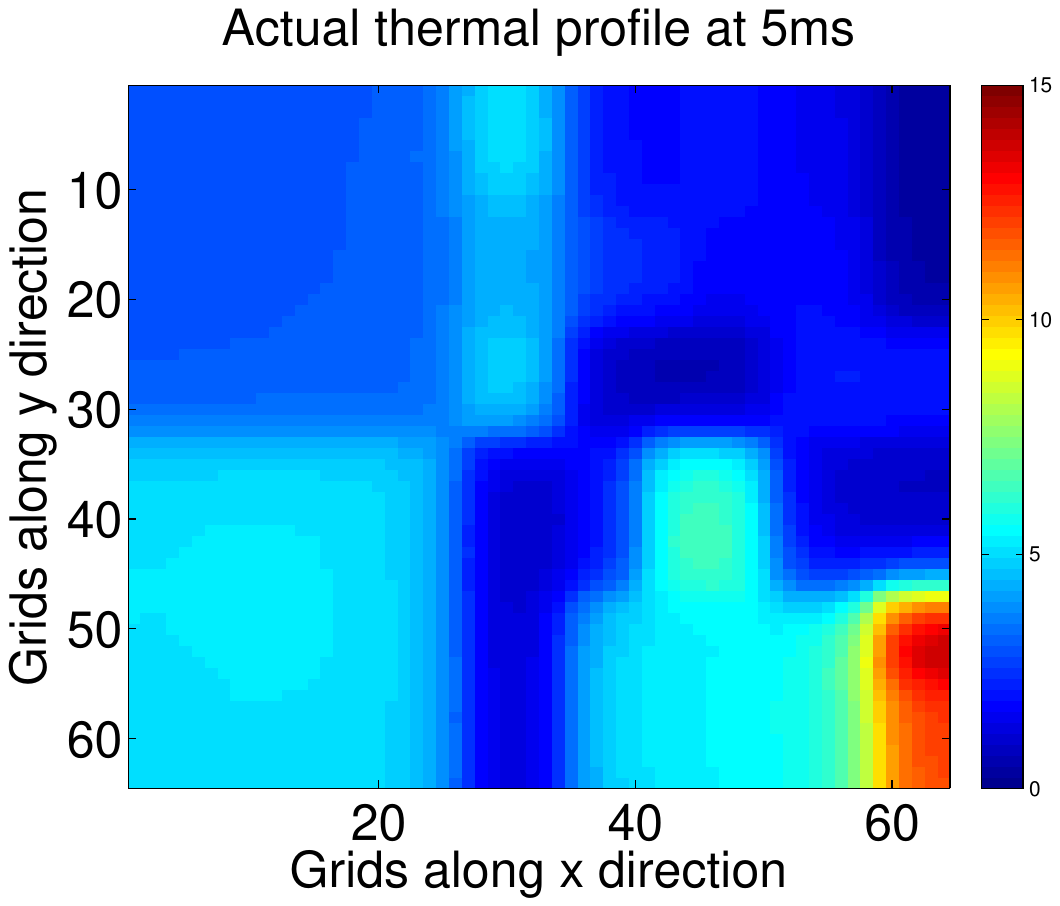}
        \caption{Temperature rise at $5~ms$}\label{fig:Ttranfullevol3}
\end{subfigure}
\caption{Power and temperature map, transient, actual \label{fig:Ttranalphaact}}
\end{figure*}

\begin{figure*}[!htbp]
                \begin{subfigure}[b]{0.24\linewidth}
	\centering
	\includegraphics[width=0.9\linewidth , trim = {.8cm 6.5cm 2cm 6.5cm}, clip=true ]{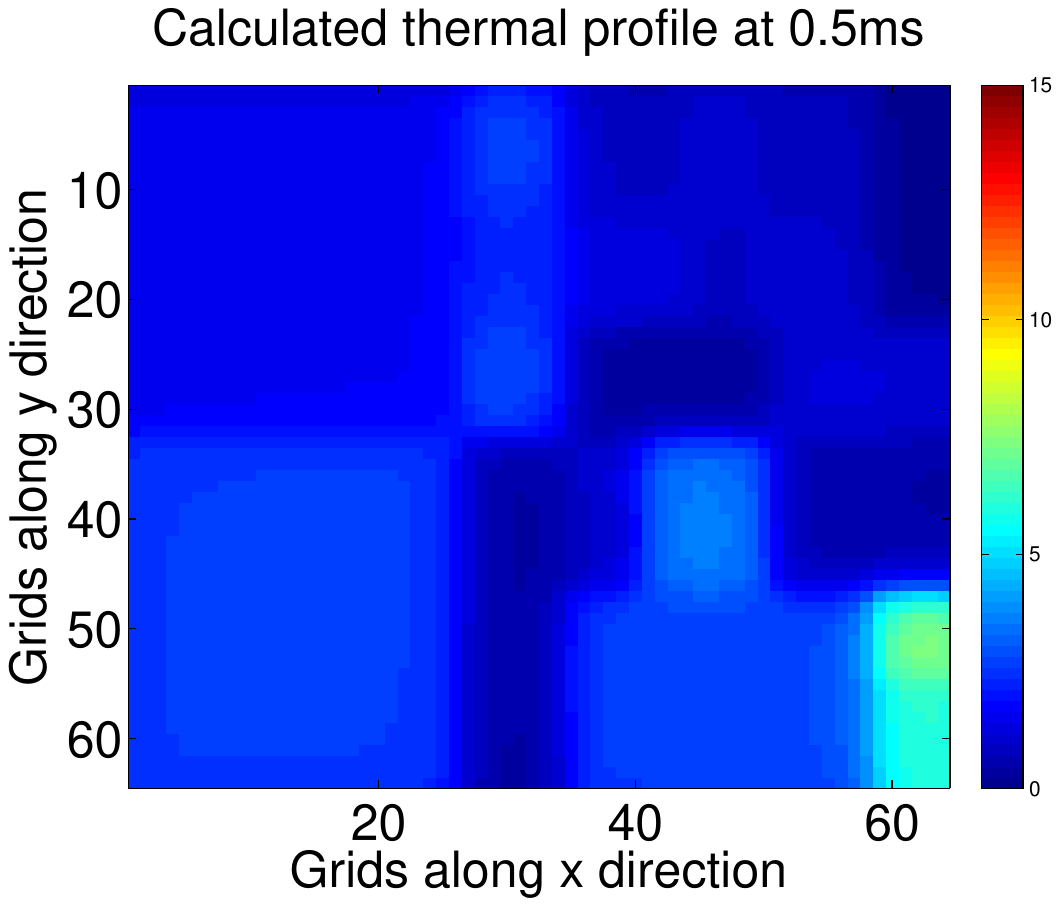}
	\caption{Temperature rise at $0.5~ms$}\label{fig:Ttranfullcalcevol0}
\end{subfigure}
                \begin{subfigure}[b]{0.24\linewidth}
	\centering
	\includegraphics[width=0.9\linewidth , trim = {.8cm 6.5cm 2cm 6.5cm}, clip=true ]{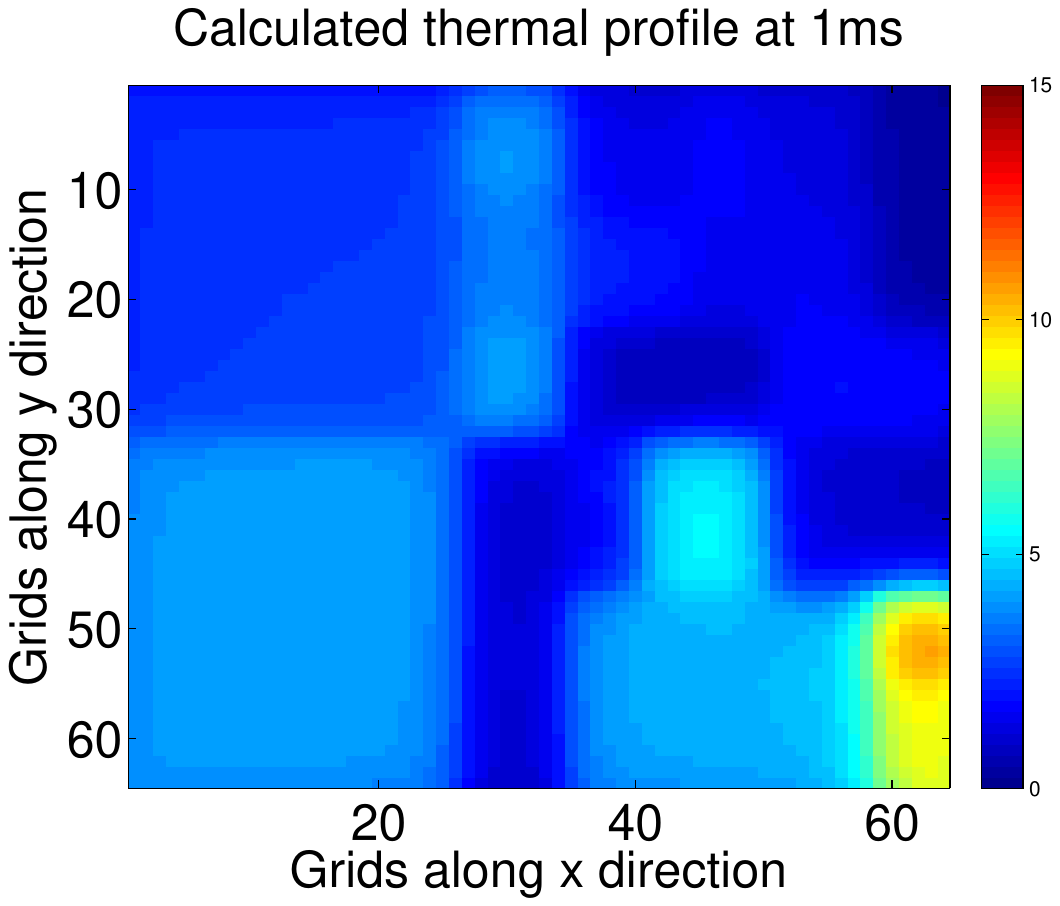}
	\caption{Temperature rise at $1~ms$}\label{fig:Ttranfullcalcevol1}
\end{subfigure}
                \begin{subfigure}[b]{0.24\linewidth}
	\centering
	\includegraphics[width=0.9\linewidth , trim = {.8cm 6.5cm 2.0cm 6.5cm}, clip=true ]{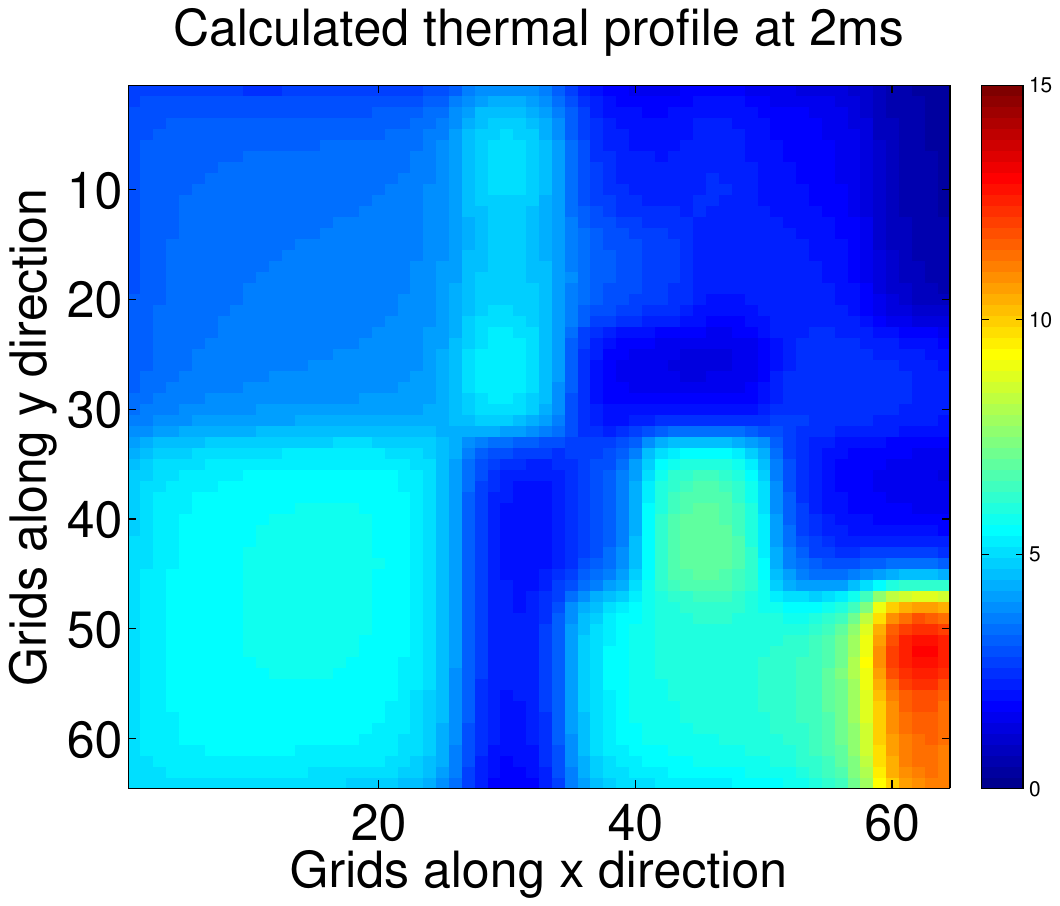}
	\caption{Temperature rise at $2~ms$}\label{fig:Ttranfullcalcevol2}
\end{subfigure}
\begin{subfigure}[b]{0.24 \linewidth}
	\centering
	\includegraphics[width=0.9\linewidth , trim = {.8cm 6.5cm 2.0cm 6.5cm}, clip=true
	]{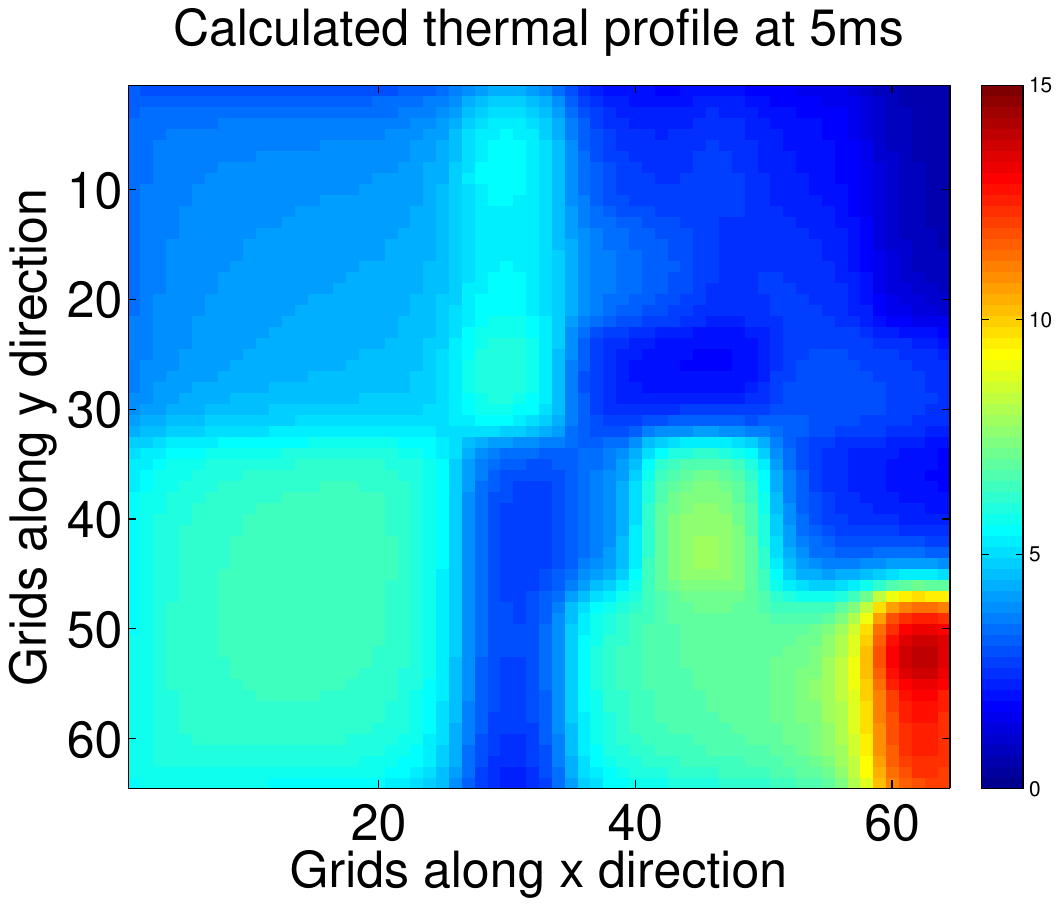}
        \caption{Temperature rise at $5~ms$}\label{fig:Ttranfullcalcevol3}
\end{subfigure}
\caption{Power and temperature map, transient, calculated \label{fig:Ttranalphacalc}}
\end{figure*}

Next, we use the derived transient Green's function to obtain the transient thermal profile for the floorplan of Alpha21264, corresponding to the power profile in Figure~\ref{fig:dynpower}.
The error at all times was observed to be less than 5\% with a simulation time of $70~ms$ for 100 time steps. 
The corresponding computed transient thermal profiles at $0.5~ms$, $1~ms$, $2~ms$, and $5~ms$ are shown in Figure~\ref{fig:Ttranalphacalc}. The corresponding actual thermal profiles obtained from a modified version of Hotspot are shown in Figure~\ref{fig:Ttranalphaact}.

\subsubsection{Full chip time-varying transient thermal profile}
\begin{figure*}

\centering

\makebox[\columnwidth]{%
\begin{minipage}[b]{0.9\columnwidth}
		\begin{center}
		\includegraphics[width=0.8\columnwidth, trim = {.8cm 6.5cm 1.6cm 6.5cm },
		clip=true]{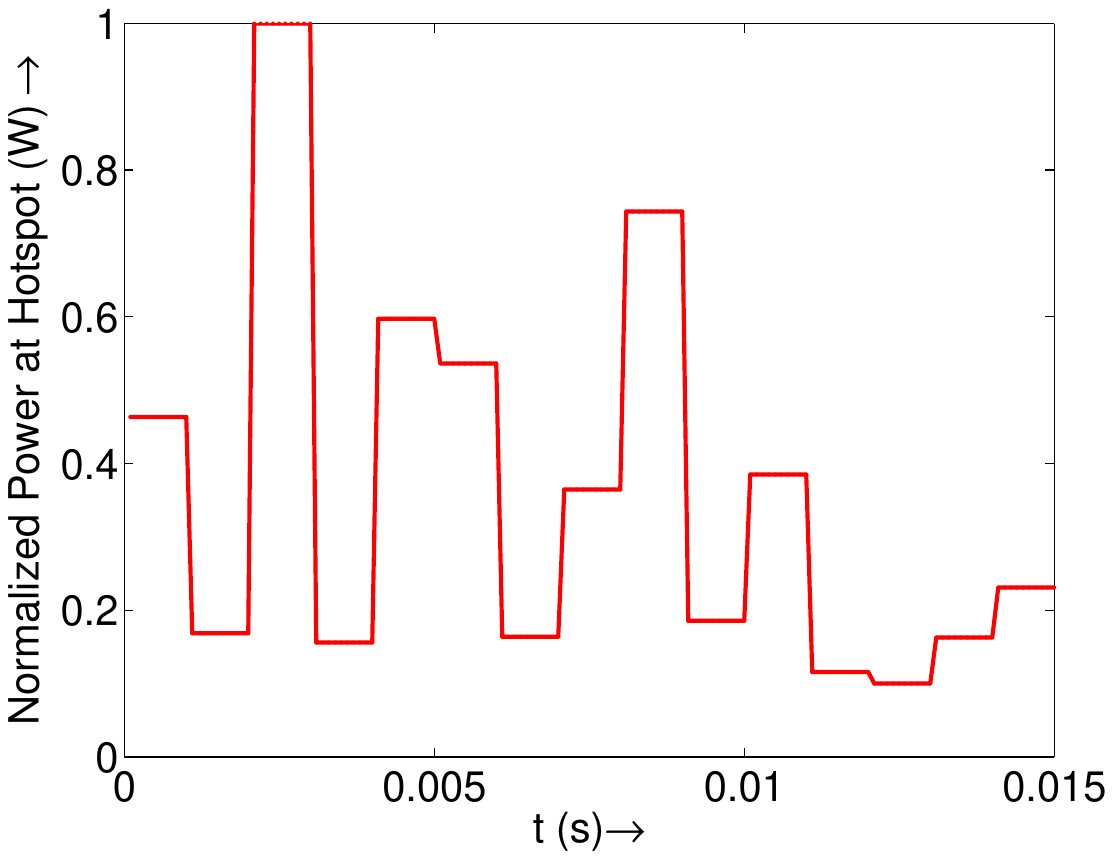}
		\end{center}
		\subcaption{Transient power variation} \label{fig:powtimevar}
	\end{minipage}%
		\begin{minipage}[b]{0.9\columnwidth}
		\begin{center}
		\includegraphics[width=0.8\columnwidth, trim = {.8cm 6.5cm 1.6cm 6.5cm },
		clip=true]{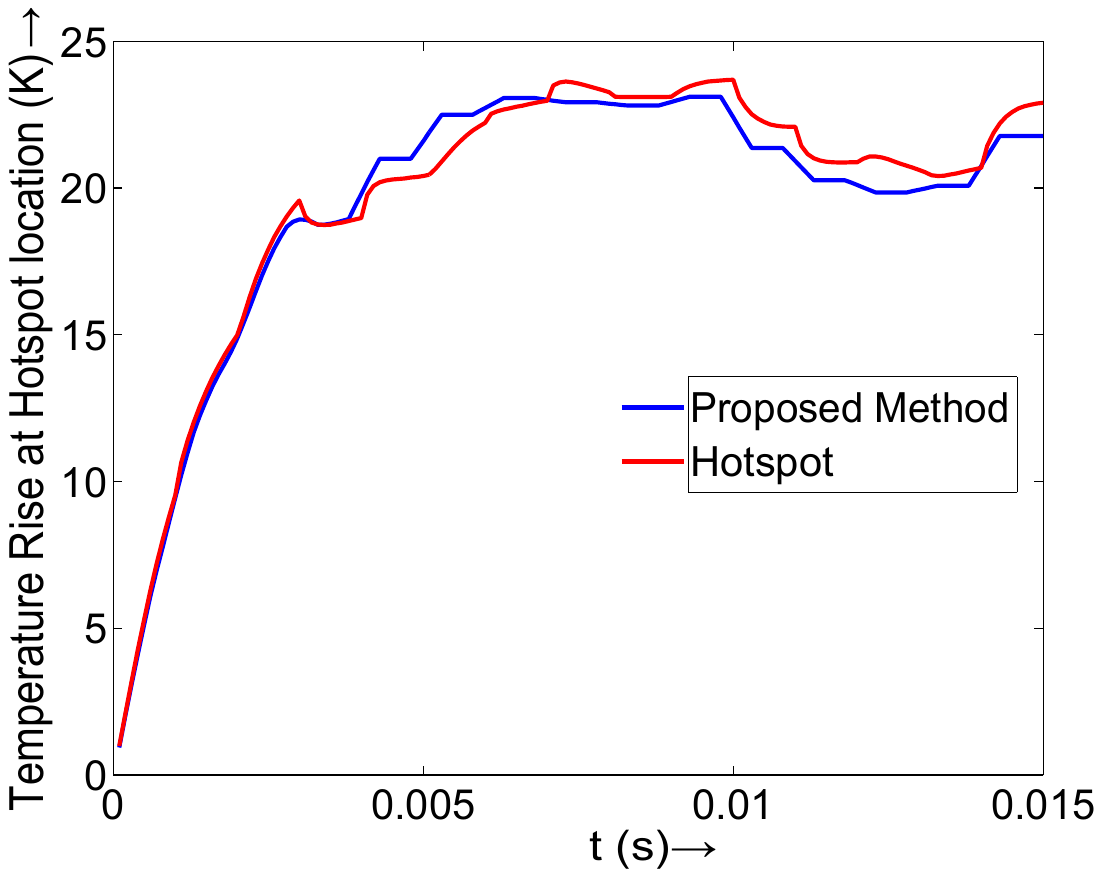}
		\end{center}
		\subcaption{Transient evolution of temperature at the hottest location} \label{fig:trantimevar}
	\end{minipage}
	}
        \caption{Transient evolution of temperature: test case 1}\label{fig:trantimevarfig}
        \end{figure*}

\noindent \textbf{Test Case 1 [Real floorplan]:} 
We randomly vary the power profile in Figure~\ref{fig:dynpower} every $1~ms$ and obtain the thermal profile for this time-varying power input using HotSpot as well as our proposed method (Equation~\ref{eqn:tranapprox}). We simulate the temperature until $15~ms$ at intervals of $0.1~ms$. Using our algorithm, we needed $290~ms$ to compute the thermal profile for 150 time steps ($\approx 2~ms$ per time step), while HotSpot took nearly 20 minutes ($4000\times$ speedup). The average error in our case was 3.8\%.
The evolution of the dynamic power profile at the hottest location is shown in Figure~\ref{fig:powtimevar}, while the corresponding thermal profile is shown in Figure~\ref{fig:trantimevar}.

\begin{figure*}

\centering        
\makebox[\columnwidth]{
\begin{minipage}[b]{0.9\columnwidth}
		\begin{center}
		\includegraphics[width=0.8\columnwidth, trim = {.8cm 6.5cm 2.0cm 6.5cm },
		clip=true]{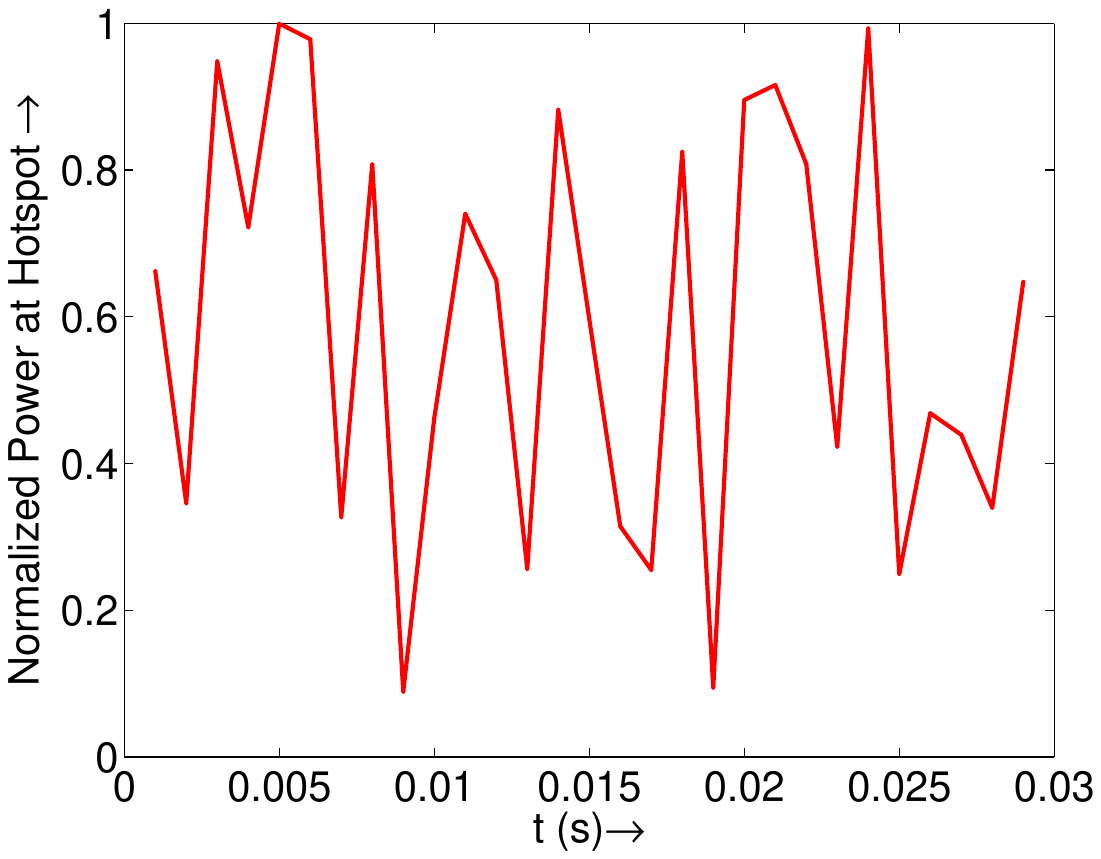}
		\end{center}
		\subcaption{Transient power variation} \label{fig:powtimevartc2}
	\end{minipage}
		\begin{minipage}[b]{0.9\columnwidth}
		\begin{center}
		\includegraphics[width=0.8\columnwidth, trim = {.8cm 6.5cm 2.0cm 6.5cm },
		clip=true]{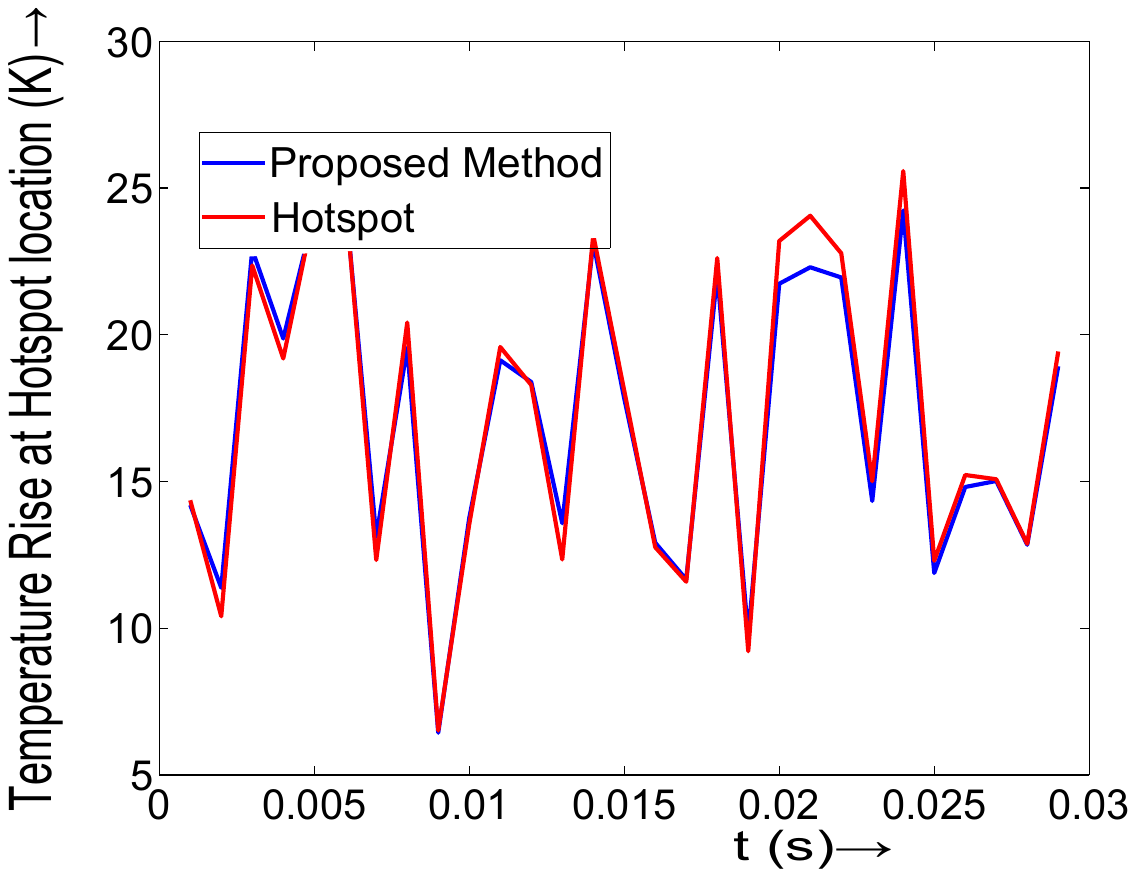}
		\end{center}
		\subcaption{Transient evolution of temperature at one location} \label{fig:trantimevartc2}
	\end{minipage}
	}
        \caption{Transient evolution of temperature: test case 2}\label{fig:trantimevarfigtc2}
        \end{figure*}
\noindent \textbf{Test Case 2 [Stress testing]:}		
Next, we randomly vary the power profile in test case 2 of the steady-state every $1~ms$ and observe the evolution of temperature over 30 time steps. The power profile at one of the locations and the corresponding thermal profiles are given in Figure~\ref{fig:trantimevarfigtc2}. Our algorithm takes $59~ms$ to compute the thermal profile for 30 time steps at intervals of $1~ms$ each ($2~ms$ per time step). In contrast, HotSpot takes over 15 minutes to compute the same thermal profile ($15000\times$ speedup). 
These results are summarized in Table~\ref{tab:exectimetran}.

\subsection{Analysis of the Results}
We have established through a wide range of test cases that our proposed method provides fast as well as accurate solutions for the effects considered. Next, we analyze the importance of modeling each of the individual effects --  temperature-dependent leakage, variation in leakage, temperature-dependent conductivity, and variation in conductivity profiles. 

To do so, we sequentially consider a subset of these effects while ignoring the rest of the effects. This helps us determine the individual thermal contribution of each of these effects in the cases we have modeled.

Table~\ref{tab:error} summarizes the results obtained in various scenarios. Figure~\ref{fig:error} graphically represents this error. We see that not accounting for any kind of variability leads to a temperature estimation error of up to 22\%. 
If we ignore the temperature dependence of conductivity but consider variability and temperature dependence 
of leakage, the error varies from 2 to 7.5\%. 
Ignoring variations in the conductivity profile leads to $<1\%$  error in thermal estimation. Thus in thermal 
modeling, the effects of random variations in the conductivity profile can be safely ignored. The need to model temperature-dependent conductivity would depend on the accuracy requirement of the design. However, considering 
variability in leakage power along with the temperature dependence of leakage power is 
\emph{absolutely essential} to achieve a meaningful simulation accuracy.

\subsubsection*{Comparison with state-of-the-art approaches}
Since there is no state-of-the-art work in thermal modeling that considers variability as well as temperature-dependent leakage, 
we compare our results against the modified version of HotSpot. Our algorithm provides a $4000-370000\times$ speedup 
over HotSpot, while maintaining the error within 4\% in all cases. Table~\ref{tab:exectime} summarizes the simulation 
speed of various tools for the steady-state.

\subsubsection*{Memory and Energy Analysis}
Since we use compact thermal models in our work, our method consumes a lot less memory compared to the existing simulators. For the steady-state, our method uses 4MB memory only, compared to 360MB for HotSpot. Similarly, in the transient case, we use approximately 85MB, compared to the 900 MB used by HotSpot. The high memory consumption in HotSpot is primarily because finite difference methods involve matrices with a very large number of nodes, whereas
Green's function methods need just as many nodes as the required temperature resolution.

Owing to its analytical nature and ultra-high speed, we outperform state-of-the-art methods in terms of energy consumption as well.

%% file: conc.tex
\section{Conclusion}
\label{sec:conc}
In this paper, we propose a fast leakage and variability-aware thermal simulation method that also captures the temperature dependence of conductivity. We derive a closed-form of the Green's function considering all these effects using novel insights and algebraic techniques. Our approach provides fast and accurate solutions for both the steady-state and the transient thermal profile and has been validated with a wide variety of test cases. As device dimensions continue to shrink, process variation has become a serious problem. The methods proposed in this work can equip designers to tackle this problem and may spawn further research in this area.

%% file: main.bbl

%% file: bios.tex
\begin{IEEEbiography}[{\includegraphics[width=1in,height=1.25in,clip,keepaspectratio]{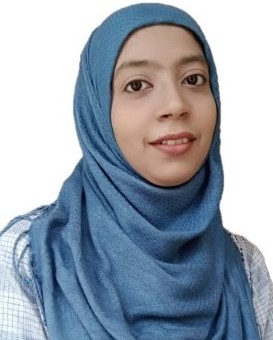}}]{Hameedah Sultan}
Hameedah Sultan has obtained her Ph.D. degree from the School of Information Technology, Indian Institute of Technology Delhi in 2021. She now works as a part of the Graphics Systems team in Qualcomm, Singapore. She has done her Masters in VLSI Design Tools and Technology, IIT Delhi. Her research interests include architectural-level power, thermal and noise modeling and its impact on performance.
\end{IEEEbiography}

\begin{IEEEbiography}[{\includegraphics[width=1in,height=1.25in,clip,keepaspectratio]{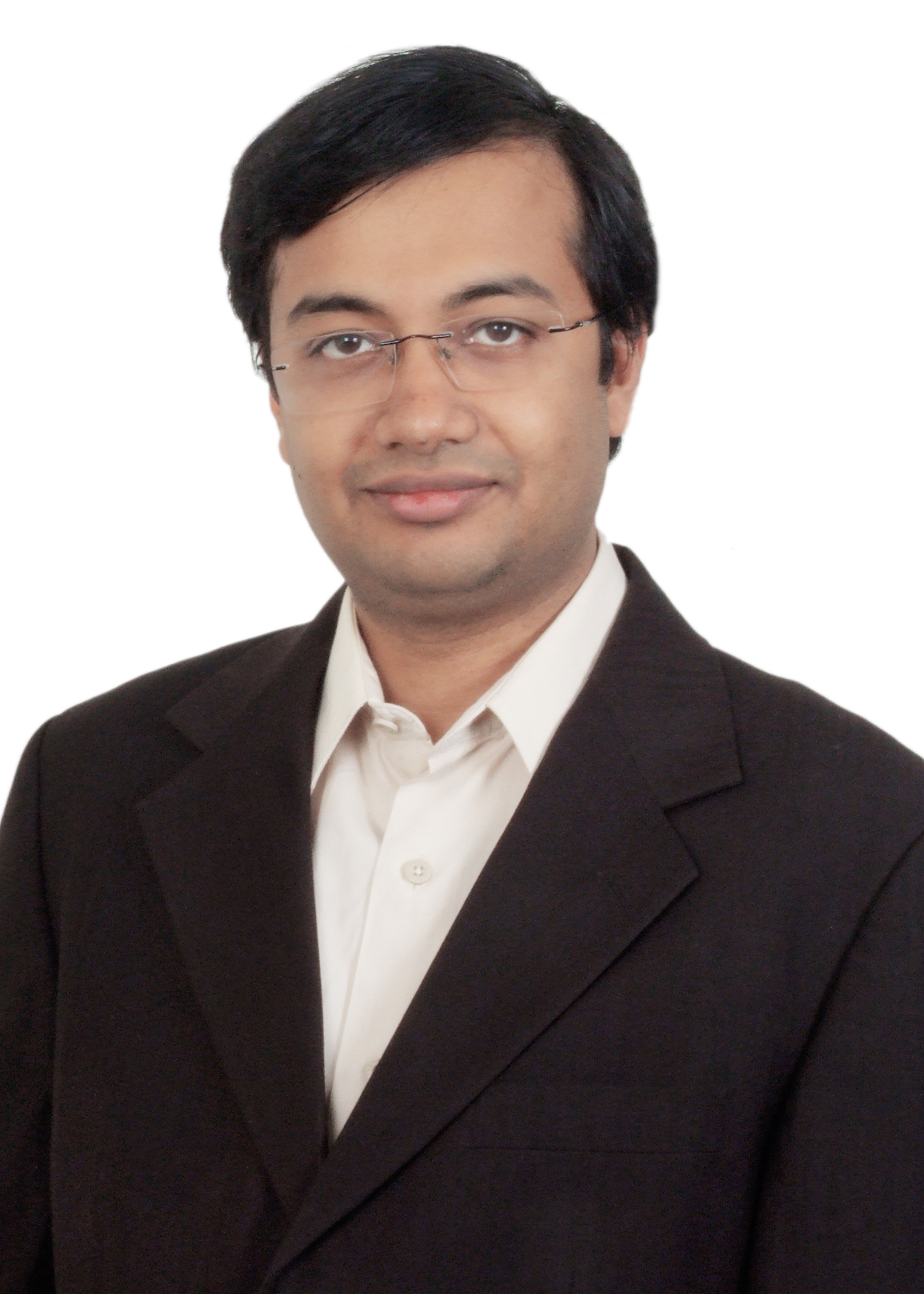}}]{Smruti R. Sarangi}
Prof. Smruti Ranjan Sarangi is an Associate Professor in the Computer Science and Engineering Department at IIT Delhi with a joint appointment in the Department of Electrical Engineering. He primarily works in parallel and distributed architectures and systems. His research areas cover multicore processors, cyber-security, emerging technologies, networks on chip, operating systems for parallel computers, and parallel algorithms. 
Dr. Sarangi obtained his Ph.D in computer architecture from the University of Illinois at Urbana Champaign(UIUC), USA  in 2006, and a B.Tech in computer science from IIT Kharagpur in 2002. He has filed five US patents, three Indian patents, and has published 87 papers in reputed international conferences and journals. He is the author of the popular undergraduate textbook on computer architecture titled, “Computer Organisation and
Architecture”, published by McGrawHill. He is a member of the IEEE and
ACM.
\end{IEEEbiography}

%% file: appendix.tex
\section{Hankel Transform and its Properties}
The Hankel transform is a mathematical transform that decomposes any function $f(r)$ into an infinite number of Bessel functions of the first kind. It is defined as:
\begin{equation}
\label{eqn:hankeldefapp}
\hankel(f(r)) = H(s) = \int_0^\infty f(r) \bessel_0(sr)r dr,
\end{equation}
where $\bessel_0$ is the Bessel function of the first kind of order 0, and $\hankel$ denotes the Hankel transform operator.

The inverse Hankel transform of $H(s)$ is defined as: 
\begin{equation}
\label{eqn:invhankeldef}
\hankel^{-1}(H(s)) = f(r) = \int_0^\infty H(s) \bessel_0(sr)s ds,
\end{equation}
where $\hankel^{-1}$ denotes the inverse Hankel transform operator.

\subsection*{Theorem I: The Hankel transform of a radially symmetric function in polar coordinates is equivalent to its 2D Fourier transform.}
\subsection*{Proof:}

The 2D Fourier transform is given by:
\begin{equation}
\fourier(f(r)) = F(u,v) = \int_{-\infty}^\infty\int_{-\infty}^\infty f(x,y) e^{-j(ux+vy)}dx dy,
\end{equation}
Let $x = rcos\theta$ and $y  = rsin\theta$.
Let $u = \rho cos\phi$ and $v = \rho sin\phi$

Substituting these in the above equation:
\begin{equation}
F(\rho, \phi) = \int_{0}^\infty\int_{-\pi}^\pi f(r,\theta)e^{-ir\rho cos(\phi - \theta)}rdrd\theta
\end{equation}

If $f(r)$ is radially symmetric, it is independent of the angle. Thus we can rewrite the above equation as:
\begin{equation}
F(\rho, \phi) = \int_{0}^\infty rf(r)dr\int_{-\pi}^\pi e^{-ir\rho cos(\phi - \theta)}d\theta
\end{equation}

Using the  deﬁnition of the zeroth-order Bessel function:
\begin{equation}
J_0(x) = \frac{1}{2\pi}\int_{-\pi}^\pi e^{-ix cos(\phi - \theta)}
\end{equation}

Using this in the equation above, we get:
\begin{equation}
F(\rho) = \fourier (f(r)) = 2\pi\int_{0}^\infty f(r)J_0(\rho r)rdr
\end{equation}
which is the same is $2\pi$ times the Hankel transform of order 0.
Thus:
\begin{equation}
F(\rho) = \fourier (f(r)) = 2\pi\hankel (f(r))
\end{equation}

\subsubsection*{Lemma I}: $G(u,v) =  \fourier(f_{sp_0}\T) = \fourier (\T) g_{sp_0}$, where $g_{sp_0} = \left( f_{sp_0} - \kappa + f_{sp_0}(0,0)\right)$
\subsection*{Proof:}

We use the zero order Hankel transform on $G(u,v)$ to reduce the 2D Fourier transform to a 1D Hankel transform. We denote the variables in polar coordinates by the $~\tilde{}~$ operator. Next, we apply integration by parts:
\begin{equation}
\begin{split}
\label{eqn:hankel1}
H(h) &= \hankel(\tilde{f}_{sp_0}\tilde{\T})
= \int_0^\infty(\tilde{f}_{sp_0}\tilde{\T})J_0(hr)r dr\\
&= \tilde{f}_{sp_0} {\int_0^\infty}\tilde{\T} J_0(hr)r dr {-} {\int_0^\infty} \tilde{f}'_{sp_0}dr  {\int_0^\infty} \tilde{\T} J_0(hr)r dr \\
&= \tilde{f}_{sp_0} \hankel(\tilde{\T}) - \int_0^\infty \tilde{f}'_{sp_0}dr \times \hankel(\tilde{\T}) \\
&=  \hankel(\tilde{\T})\left(\tilde{f}_{sp_0} - \int_0^\infty \frac{\partial{\tilde{f}_{sp_0}}}{\partial{r}} dr \right)\\
&=  \hankel(\tilde{\T})\left(\tilde{f}_{sp_0} - \tilde{f}_{sp_0}(\infty) + \tilde{f}_{sp_0}(0) \right)
\end{split}
\end{equation}
where $'$ denotes the derivative with respect to $r$.

Let $\tilde{f}_{sp_0}(\infty) = \kappa$.
The equivalent expression in the Cartesian coordinates becomes:
\begin{equation}
\begin{split}
\label{eqn:leakvar2}
G(u,v) 
	   &= \fourier(f_{sp_0}\T) = \fourier (\T)\underbrace{\left( f_{sp_0} - \kappa + f_{sp_0}(0,0)\right)}_{= g_{sp_0}} \\
	   &=\fourier (\T) g_{sp_0}
\end{split}
\end{equation}
{where $\fourier$ is the Fourier transform operator.}

%% file: main.bbl
\begin{thebibliography}{10}
\providecommand{\url}[1]{#1}
\csname url@samestyle\endcsname
\providecommand{\newblock}{\relax}
\providecommand{\bibinfo}[2]{#2}
\providecommand{\BIBentrySTDinterwordspacing}{\spaceskip=0pt\relax}
\providecommand{\BIBentryALTinterwordstretchfactor}{4}
\providecommand{\BIBentryALTinterwordspacing}{\spaceskip=\fontdimen2\font plus
\BIBentryALTinterwordstretchfactor\fontdimen3\font minus
  \fontdimen4\font\relax}
\providecommand{\BIBforeignlanguage}[2]{{%
\expandafter\ifx\csname l@#1\endcsname\relax
\typeout{** WARNING: IEEEtran.bst: No hyphenation pattern has been}%
\typeout{** loaded for the language `#1'. Using the pattern for}%
\typeout{** the default language instead.}%
\else
\language=\csname l@#1\endcsname
\fi
#2}}
\providecommand{\BIBdecl}{\relax}
\BIBdecl

\bibitem{elseviermcm}
``Thermal layout optimization for 3d stacked multichip modules,''
  \emph{Microelectronics Journal}, vol. 139, p. 105882, 2023.

\bibitem{sapatnekarvar}
Y.~Zhan, B.~Goplen, and S.~Sapatnekar, ``Electrothermal analysis and
  optimization techniques for nanoscale integrated circuits,'' in
  \emph{ASPDAC'06}.

\bibitem{isac}
Y.~Yang, Z.~Gu, C.~Zhu, R.~P. Dick, and L.~Shang, ``{ISAC}: Integrated
  space-and-time-adaptive chip-package thermal analysis,'' \emph{IEEE TCAD},
  vol.~26, no.~1, pp. 86--99, 2006.

\bibitem{3dsim}
H.~Sultan and S.~Sarangi, ``A fast leakage aware thermal simulator for {3D}
  chips,'' in \emph{DATE'17}.

\bibitem{sapatnekar}
Y.~Zhan and S.~S. Sapatnekar, ``High-efficiency green function-based thermal
  simulation algorithms,'' \emph{Computer-Aided Design of Integrated Circuits
  and Systems, IEEE Transactions on}, vol.~26, no.~9, pp. 1661--1675, 2007.

\bibitem{fourier}
D.~V. Widder, \emph{The heat equation}.\hskip 1em plus 0.5em minus 0.4em\relax
  Academic Press, 1976, vol.~67.

\bibitem{greenintro}
J.~V. Beck, K.~D. Cole, A.~Haji-Sheikh, and B.~Litkouhl, \emph{Heat conduction
  using Green's function}.\hskip 1em plus 0.5em minus 0.4em\relax Taylor \&
  Francis, 1992.

\bibitem{lightsim}
S.~Sarangi, G.~Ananthanarayanan, and M.~Balakrishnan, ``Lightsim: A leakage
  aware ultrafast temperature simulator,'' in \emph{ASPDAC'14}.

\bibitem{powerblur2014}
A.~Ziabari, J.-H. Park, E.~K. Ardestani, J.~Renau, S.-M. Kang, and A.~Shakouri,
  ``Power blurring: Fast static and transient thermal analysis method for
  packaged integrated circuits and power devices,'' \emph{IEEE TVLSI}, vol.~22,
  no.~11, pp. 2366--2379, 2014.

\bibitem{mittal}
S.~Mittal, ``A survey of architectural techniques for managing process
  variation,'' \emph{CSUR}, vol.~48, no.~4, p.~54, 2016.

\bibitem{varius}
S.~Sarangi, B.~Greskamp, R.~Teodorescu, J.~Nakano, A.~Tiwari, and J.~Torrellas,
  ``Varius: A model of process variation and resulting timing errors for
  microarchitects,'' \emph{IEEE TSM}, vol.~21, no.~1, pp. 3--13, 2008.

\bibitem{skadronvar}
E.~Humenay, D.~Tarjan, and K.~Skadron, ``Impact of process variations on
  multicore performance symmetry,'' in \emph{DATE 2007}.

\bibitem{liu}
Y.~Liu, R.~P. Dick, L.~Shang, and H.~Yang, ``Accurate temperature-dependent
  integrated circuit leakage power estimation is easy,'' in \emph{DATE 2007}.

\bibitem{tempsurvey}
H.~Sultan, A.~Chauhan, and S.~R. Sarangi, ``A survey of chip-level thermal
  simulators,'' \emph{CSUR}, vol.~52, no.~2, pp. 1--35, 2019.

\bibitem{leungvar}
G.~Leung and C.~O. Chui, ``Variability impact of random dopant fluctuation on
  nanoscale junctionless {FinFETs},'' \emph{IEEE Electron Device Letters},
  vol.~33, no.~6, pp. 767--769, 2012.

\bibitem{burzo}
M.~G. Burzo, P.~L. Komarov, and P.~Raad, ``Non-contact thermal conductivity
  measurements of p-doped and n-doped gold covered natural and
  isotopically-pure silicon and their oxides,'' in \emph{EuroSimE'04}.

\bibitem{hotspot}
W.~Huang, S.~Ghosh, S.~Velusamy, K.~Sankaranarayanan, K.~Skadron, and M.~R.
  Stan, ``Hotspot: A compact thermal modeling methodology for early-stage
  {VLSI} design,'' \emph{VLSI Systems, IEEE Transactions on}, vol.~14, no.~5,
  pp. 501--513, 2006.

\bibitem{pod}
L.~Jiang, A.~Dowling, Y.~Liu, and M.-C. Cheng, ``Chip-level thermal simulation
  for a multicore processor using a multi-block model enabled by proper
  orthogonal decomposition,'' in \emph{iTherm, 2022.}

\bibitem{systemc}
Y.~Chen, S.~Vinco, E.~Macii, and M.~Poncino, ``Systemc-ams thermal modeling for
  the co-simulation of functional and extra-functional properties,'' \emph{ACM
  Transactions on Design Automation of Electronic Systems (TODAES)}, vol.~24,
  no.~1, pp. 1--26, 2018.

\bibitem{deepoheat}
Z.~Liu, Y.~Li, J.~Hu, X.~Yu, S.~Shiau, X.~Ai, Z.~Zeng, and Z.~Zhang,
  ``Deepoheat: Operator learning-based ultra-fast thermal simulation in 3d-ic
  design.''

\bibitem{pathania3dttp}
S.~Niknam, Y.~Shen, A.~Pathania, and A.~D. Pimentel, ``3d-ttp: Efficient
  transient temperature-aware power budgeting for 3d-stacked processor-memory
  systems.''

\bibitem{coskun2023}
P.~Shukla, V.~F. Pavlidis, E.~Salman, and A.~K. Coskun, ``Tread-m3d:
  Temperature-aware dnn accelerators for monolithic 3d mobile systems,''
  \emph{IEEE Transactions on Computer-Aided Design of Integrated Circuits and
  Systems}, 2023.

\bibitem{jaffari}
J.~Jaffari and M.~Anis, ``Statistical thermal profile considering process
  variations: Analysis and applications,'' \emph{IEEE TCAD}, vol.~27, no.~6,
  pp. 1027--1040, 2008.

\bibitem{varipower}
K.~Meng, F.~Huebbers, R.~Joseph, and Y.~Ismail, ``Modeling and characterizing
  power variability in multicore architectures,'' in \emph{ISPASS'07}.

\bibitem{juan}
D.-C. Juan, S.~Garg, and D.~Marculescu, ``Statistical thermal evaluation and
  mitigation techniques for {3D} chip-multiprocessors in the presence of
  process variations,'' in \emph{DATE'11}.

\bibitem{henkelvar}
M.~Shafique, D.~Gnad, S.~Garg, and J.~Henkel, ``Variability-aware dark silicon
  management in on-chip many-core systems,'' in \emph{DATE 2015}, 2015, pp.
  387--392.

\bibitem{pvsmartphone}
G.~Prasad~Srinivasa, S.~Haseley, G.~Challen, and M.~Hempstead, ``Quantifying
  process variations and its impacts on smartphones,'' in \emph{ISPASS, 2019.}

\bibitem{isacvar}
B.~Li, L.-S. Peh, and P.~Patra, ``Impact of process and temperature variations
  on network-on-chip design exploration,'' in \emph{NOCS)}, 2008, pp. 117--126.

\bibitem{hotspot5}
W.~Huang, K.~Skadron, S.~Gurumurthi, R.~J. Ribando, and M.~R. Stan,
  ``Differentiating the roles of {IR} measurement and simulation for power and
  temperature-aware design,'' in \emph{ISPASS, 2009.}

\bibitem{ziabari}
A.~Ziabari, Z.~Bian, and A.~Shakouri, ``Adaptive power blurring techniques to
  calculate {IC} temperature profile under large temperature variations,''
  \emph{IMAPS'10}.

\bibitem{insulators}
a.~Köroğlu and E.~Pop, ``High thermal conductivity insulators for thermal
  management in 3d integrated circuits,'' \emph{IEEE Electron Device Letters},
  pp. 1--1, 2023.

\bibitem{MLtran}
A.~Kumar, N.~Chang, D.~Geb, H.~He, S.~Pan, J.~Wen, S.~Asgari, M.~Abarham, and
  C.~Ortiz, ``Ml-based fast on-chip transient thermal simulation for
  heterogeneous 2.5d/3d ic designs,'' in \emph{VLSI-DAT, 2022.}, 2022, pp.
  1--8.

\bibitem{comet}
L.~Siddhu, R.~Kedia, S.~Pandey, M.~Rapp, A.~Pathania, J.~Henkel, and P.~R.
  Panda, ``Comet: An integrated interval thermal simulation toolchain for 2d,
  2.5d, and 3d processor-memory systems,'' 2022.

\bibitem{he}
Z.~He, W.~Cui, C.~Cui, T.~Sherwood, and Z.~Zhang, ``Efficient uncertainty
  modeling for system design via mixed integer programming,'' in \emph{ICCAD,
  2019.}, 2019, pp. 1--8.

\bibitem{pvnoc}
s.~v.~r. chittamuru, I.~G. Thakkar, and S.~Pasricha, ``Libra: Thermal and
  process variation aware reliability management in photonic
  networks-on-chip,'' \emph{IEEE Transactions on Multi-Scale Computing
  Systems}, vol.~4, no.~4, pp. 758--772, 2018.

\bibitem{varsim}
H.~Sultan and S.~R. Sarangi, ``Varsim: a fast and accurate variability and
  leakage aware thermal simulator,'' in \emph{DAC 2020}.\hskip 1em plus 0.5em
  minus 0.4em\relax IEEE, 2020, pp. 1--6.

\bibitem{yuK}
Z.~Yu, D.~Yergeau, and R.~W. Dutton, ``Full chip thermal simulation,'' in
  \emph{ISQED'00}.

\bibitem{hotspot6}
R.~Zhang, M.~R. Stan, and K.~Skadron, ``Hotspot 6.0: Validation, acceleration
  and extension,'' University of Virginia, Tech. Rep., 2015.

\end{thebibliography}
